\documentclass[iop]{emulateapj}

\usepackage{amsmath}
\usepackage{amssymb}
\usepackage[none]{hyphenat}
\usepackage{natbib}
\usepackage{units}
\usepackage{color}
\usepackage{textcomp}
\usepackage{placeins}
\usepackage{bm}

\newcommand{\RN}[1]{\textup{\uppercase\expandafter{\romannumeral#1}}}

\newcommand\TopStrut{\rule{0pt}{2.6ex}}       
\newcommand\BottomStrut{\rule[-1.2ex]{0pt}{0pt}} 

\shorttitle{Stellar orbits in the Galactic Center}
\shortauthors{Gillessen et al.}

\begin{document}

\title{An Update on Monitoring Stellar Orbits in the Galactic Center}
\author{S.~Gillessen\altaffilmark{1}, P.M.~Plewa\altaffilmark{1}, F.~Eisenhauer\altaffilmark{1}, R.~Sari\altaffilmark{2}, I.~Waisberg\altaffilmark{1}, M.~Habibi\altaffilmark{1}, O.~Pfuhl\altaffilmark{1}, E.~George\altaffilmark{1}, J.~Dexter\altaffilmark{1}, S.~von~Fellenberg\altaffilmark{1}, T.~Ott\altaffilmark{1}, R.~Genzel\altaffilmark{1,3} }
\altaffiltext{1}{Max-Planck-Institut f\"ur Extraterrestrische Physik, 85748 Garching, Germany}
\altaffiltext{2}{Racah Institute of physics, The Hebrew University, Jerusalem 91904, Israell}
\altaffiltext{3}{Astronomy \& Physics Departments, University of California, Berkeley, CA 94720, USA}

\begin{abstract}
Using 25 years of data from uninterrupted monitoring of stellar orbits in the Galactic Center, we present an update of the main results from this unique data set: A measurement of mass of and distance to Sgr~A*. Our progress is not only due to the eight year increase in time base, but also due to the improved definition of the coordinate system. 
The star S2 continues to yield the best constraints on the mass of and distance to Sgr~A*; the statistical errors of $0.13\times10^6\,M_\odot$ and $0.12\,$kpc have halved compared to the previous study. 
The S2 orbit fit is robust and does not need any prior information. Using coordinate system priors, also the star S1 yields tight constraints on mass and distance. 
For a combined orbit fit, we use 17 stars, which yields our current best estimates for mass and distance: $M = 4.28 \pm 0.10|_\mathrm{stat.} \pm 0.21|_\mathrm{sys} \times 10^6 \, M_\odot$ and $R_0 = 8.32 \pm 0.07|_\mathrm{stat.} \pm 0.14|_\mathrm{sys} \, \mathrm{kpc}$. 
These numbers are in agreement with the recent determination of $R_0$ from the statistical cluster parallax. The positions of the mass, of the near-infrared flares from Sgr~A* and of the radio source Sgr~A* agree to within $1\,$mas. In total, we have determined orbits for 40 stars so far, a sample which consists of 32 stars with randomly oriented orbits and a thermal eccentricity distribution, plus eight stars for which we can explicitly show that they are members of the clockwise disk of young stars, and which have lower eccentricity orbits.
\end{abstract}

\section{Introduction}

The near-infrared regime is a sweet spot for studying the gravitational potential in the Galactic Center. For measuring the latter, one would like to have as high a resolution as possible, and have access to the emission of objects compact and bright enough that they can serve as test particles for the potential. The optimum band is around $2\,\mu$m wavelength, where the extinction screen amounts to less than $3\,$mag \citep{2011ApJ...737...73F}, and where adaptive optics at $8\,$m-class telescopes is performing well for typical atmospheric conditions. The intrinsic resolution of around $50\,$mas allows measuring stellar orbits with semi-major axes of similar size, corresponding to orbital periods around a decade for a black hole of $4\,$million solar masses in $8\,$kpc distance.

25 years of near-infrared observations of the Galactic Center have shown that a wealth of fundamental astrophysical and physical questions can be addressed with these measurements, ranging from star formation, to stellar dynamics, to testing general relativity \citep{2010RvMP...82.3121G}. The outstanding, main result is that the radio source Sgr~A* is a massive black hole in the center of the Milky Way. The key to this result is that one can measure Sgr~A*'s mass from tracing individual stellar orbits around it. If a sufficiently large part of an orbit is sampled, one can deduce information on the potential through which the star is moving. In particular, one can determine the central mass and the distance to it. Due to its proximity, the Galactic Center is the only galactic nucleus where such an experiment is currently feasible. 

A geometric determination of the distance to the Galactic Center, $R_0$, is important for many branches of astronomy. $R_0$ is one of the fundamental parameters of any model of the Milky Way, and its value determines mass and size of the galaxy. This ties $R_0$ into the cosmological distance ladder, since galactic variables serve as zero point for the period-luminosity relations determined usually in the Large Magellanic Cloud. The mass of the massive black hole is equally important. Using this value, one can place the Milky Way onto scaling relations \citep{2013ARA&A..51..511K}. Knowing the mass of and distance to Sgr~A* is the reason why the Galactic Center is a unique testbed concerning massive black holes and their vicinities for numerous models in many branches of astrophysics \citep{2014ARA&A..52..529Y}.   
The Milky Way also serves as a check for mass measurements in other galaxies, since the true black hole mass is known and one can simulate observations at lower resolution \citep{2014A&A...570A...2F}.

The most profound result of the orbital work is the proof of existence of astrophysical massive black holes. This opens up a new route of testing general relativity, at a mass scale and a field curvature that have not been accessible so far. Since the fundamental parameters are known for Sgr~A*, one can think of more ambitious experiments using the black hole. Most notably, in the near future two observations might become feasible: (i) Using the motions of stars and/or that of plasma radiating very close to the event horizon, one might be able to measure the spin of the black hole. The instrumental route to that goal is near-infrared interferometry \citep{2011Msngr.143...16E}. (ii) A global radio-interferometric array operating at around $1\,$mm should be able to resolve Sgr~A*, i.e. to deliver an actual image of the black hole's shadow \citep{1979A&A....75..228L, 2000ApJ...528L..13F, 2008Natur.455...78D}. Additionally, the most stringent tests of general relativity would be possible, if a pulsar representing a perfect clock in a short-period orbit around Sgr~A* were found \citep{2015arXiv151000394P}.

Here, we report on updates to our ongoing, long-term program of monitoring of stellar orbits around Sgr~A*. The first orbit determination dates back to 2002 \citep{2002Natur.419..694S, 2003ApJ...586L.127G}. A few years later, orbits for a handful of stars had been determined \citep{2005ApJ...620..744G, 2005ApJ...628..246E}, and again a few years later, the number of known orbits exceeded 20 \citep{2009ApJ...692.1075G}. To date, the number of known orbits has risen to around 40, and both statistical and systematic errors of our measurements are much reduced compared to previous measurements.

\section{Data}

This work is an update and improvement over our previous orbital study \citep{2009ApJ...692.1075G}. The two main improvements are:
\begin{itemize}
\item Since the previous work we have added eight more years of data, using the adaptive optics (AO) imager NACO on the VLT \citep{1998SPIE.3354..606L, 1998SPIE.3353..508R} and the AO-assisted integral field spectrograph SINFONI \citep{2003SPIE.4841.1548E, 2003SPIE.4839..329B}. This extends our time base from 17 to 25 years for the imaging and from five to 13 years for the spectroscopy. We add (in the best cases) 78 epochs of imaging and 19 epochs of spectroscopy.
\item We implement the improved reference frame described in \cite{2015MNRAS.453.3234P}. This greatly improves our prior knowledge, where we expect an orbit fit to reconstruct the mass responsible for the orbital motion of the S-stars. The new calibration links to the International Celestial Reference Frame (ICRF) in a two-step procedure, and compared to our previous work does not rely on the assumption that the mean motion of a large sample of stars observed around Sgr~A* is zero.
\end{itemize}

The other steps of the analysis are identical, and we refer the reader to \cite{2009ApJ...692.1075G} for more details. In particular, the following critical issues are treated identical:
\begin{itemize}
\item The assignment of statistical errors to individual data points. 
\item The relative weight between the earlier Speckle data (1992-2001, \cite{1992ESOC...42..617H}) and the AO data is unchanged. We weigh down the NACO-based astrometric data by a global factor 1.42, determined in \cite{2009ApJ...692.1075G} as the factor which makes the noise in the AO data match the statistical error estimates. 
\item The errors assigned to S2 in 2002 are identical to the ones derived in the previous work. S2 may have been confused and its position perturbed in 2002, when the star passed the pericenter of its orbit.
\end{itemize}

Our imaging data set contains an interruption. During 2014 and in spring 2015, NACO was not available at the VLT, resulting in significant gaps in our time series. In summer 2015 NACO resumed operation (now at UT1, no longer at UT4). The data obtained after that (ten epochs) show that we can reconstruct the stellar positions to the same level of precision as before. Also, there is no systematic mismatch between the positions obtained before and after. We therefore do not need to apply any corrections related to the interruption. 

For orbits fits using a combined data set of both VLT- and Keck-based observations, we replace the data points of \cite{2008ApJ...689.1044G} by the newer publication of the same team from \cite{2016ApJ...830...17B}.

\begin{figure}
\centering
\includegraphics[width=0.95\linewidth]{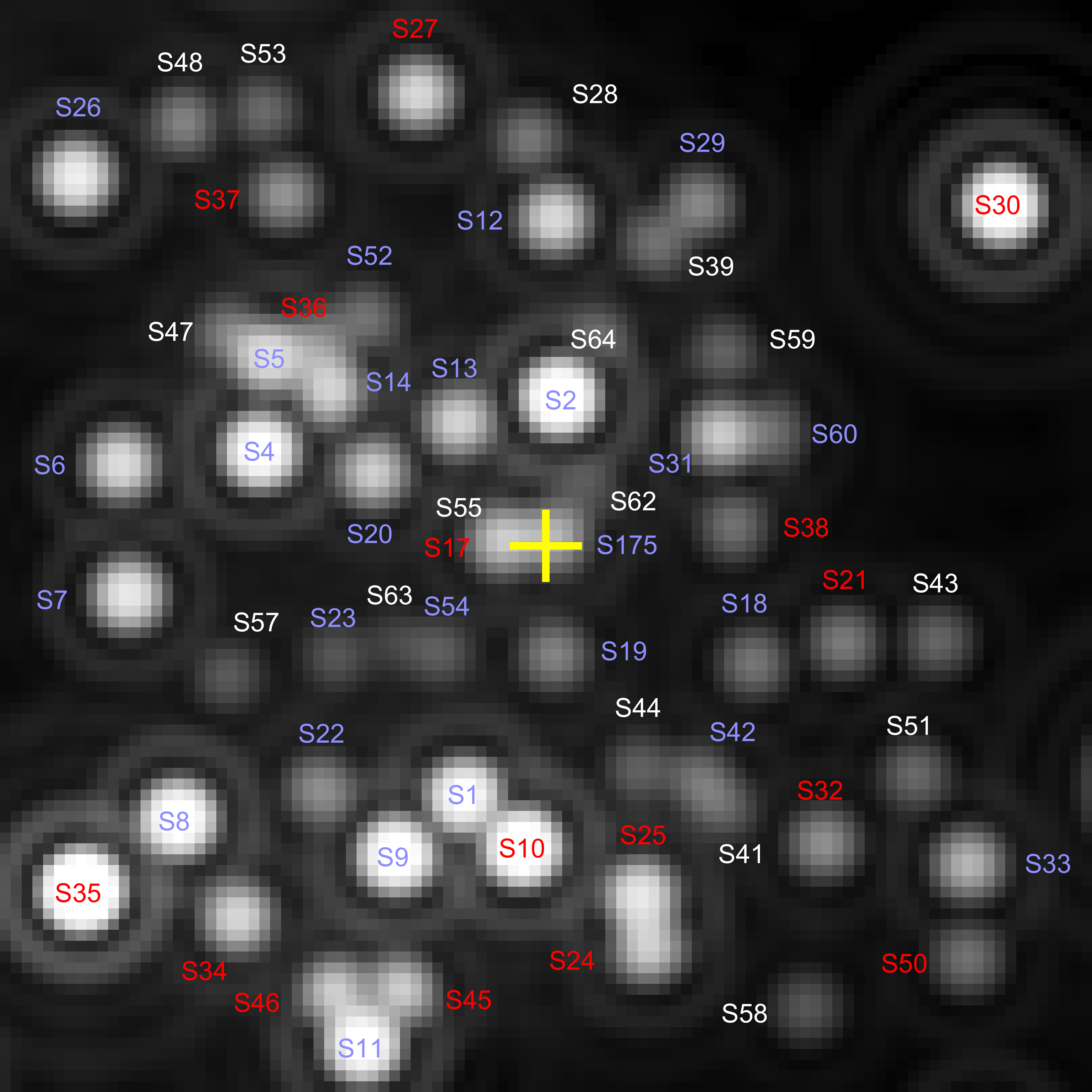}
\caption{Mock image of the central arcsecond for our reference epoch 2009.0, constructed from the measured motions and magnitudes of the stars, assuming a PSF size and pixel sampling as in our NACO data. Stars with spectral identification have colored labels, blue for early-type stars (Br-$\gamma$ absorption line detected), red for late-type stars (CO band heads detected). The yellow cross denotes the position of Sgr~A*.}
\label{fig_mockImage}
\end{figure}

\section{The gravitational potential in the Galactic Center}

\subsection{Orbit fitting}

Orbit fitting has a relatively large number of free parameters. While in its most simple form the potential has only one free physical parameter (the central mass $M$), we do not know a priori where the mass is located and how it moves. Hence six additional parameters need to be determined simultaneously: The distance to the mass $R_0$, its position on sky $(\alpha, \delta)$ and to a very good first order approximation its motion ($v_\alpha, v_\delta, v_z$). Furthermore, the orbit of the star that is probing the potential needs to be determined at the same time. The orbit parameters are essentially the initial conditions for its motion in the potential (three position variables and three velocity variables), which conventionally are expressed in terms of the classical orbital elements $(a, e, i, \Omega, \omega, t_P)$\footnote{Here, $a$ is the semi-major axis of the orbit, $e$ the eccentricity, $i$ the inclination, $\Omega$ the position angle of the ascending node, $\omega$ the longitude of the pericenter and $t_P$ the epoch of pericenter passage.}. Therefore the most simple orbit fit has already 13 free parameters. For n stars, one has $7+6 \times n$ free parameters. 

For the six coordinate system parameters of the potential we have prior knowledge. $R_0$ has been determined in multiple ways; for a recent, robust example see \cite{2014ApJ...783..130R} who report $R_0 = 8.34 \pm 0.16\,$kpc. As $R_0$ is one of the parameters we wish to determine, we do not use the prior information for it.

Since we correct our radial velocity measurements to the local standard of rest (LSR), and the expected motion of Sgr~A* is very small (below $1\,$km/s, \cite{2004ApJ...616..872R}) we expect $v_z = 0$, with an uncertainty due to the uncertainty of the LSR  of around $5\,$km/s. For the position on sky and the motion in the plane of the sky of Sgr~A*, we can adopt the limits from \cite{2015MNRAS.453.3234P}: $(\alpha, \delta) = (0,0) \pm (0.2, 0.2)\,$mas at our reference epoch 2009.0, and $(v_\alpha, v_\delta) = (0,0) \pm (0.1, 0.1)\,$mas/yr. These priors only show very small covariances, which we neglect in our analysis. 

We use a fitting code we developed in Mathematica \citep{math}, which calculates the positions and velocities by explicitly integrating the orbits, and which then uses the built-in minimization routines to find the best fitting parameters. Our code also allows for using (or omitting) prior information for the parameters to be solved for. We can either obtain the parameter errors from the inverse of the correlation matrix or from the parameter distributions as output by a Markov chain. Our fitting routine also allows fitting for more complicated orbit models, and in particular we can include relativistic corrections. We have implemented the potential of the Schwarzschild metric \citep{2008ApJ...674L..25W}, the gravitational redshift, the transverse Doppler effect and the Roemer time delay \citep{2006ApJ...639L..21Z}. Also, it is possible to integrate the orbits in the potential of an extended mass distribution.

We explicitly tested our implementation of the Schwarzschild precession term. We fitted a simulated, relativistic orbit of the star S2 (i.e. in the weak field limit) for a full orbital period (from $t = 0$ to $t = T$), which yielded back the parameters put into the simulation. Simulating the same orbit from $t = T$ to $t = 2T$ and fitting it yielded back the same parameters again, except for $\omega$ (the longitude of periastron describing the orientation of the ellipse in its plane) that had changed by the amount expected from the formula
\begin{equation}
\Delta \omega = 6\pi \frac{G\,M}{c^2}\frac{1}{a(1-e^2)}\,\,.
\end{equation}

\subsection{The potential based on S2 only}

Most of our knowledge of the MBH's potential is due to a single star, S2, which happens to be comparably bright ($m_K \approx 14$) and orbits Sgr~A* on a short period orbit ($P=16\,$yr). It is the brightest star for which we can determine an orbit and hence less prone to errors due to confusion than all other stars. Its orbital period of around 16 years is the second shortest known. We investigate the potential we can derive from S2 alone before including data from more stars.

\begin{figure*}
\centering
\includegraphics[width=0.49\linewidth]{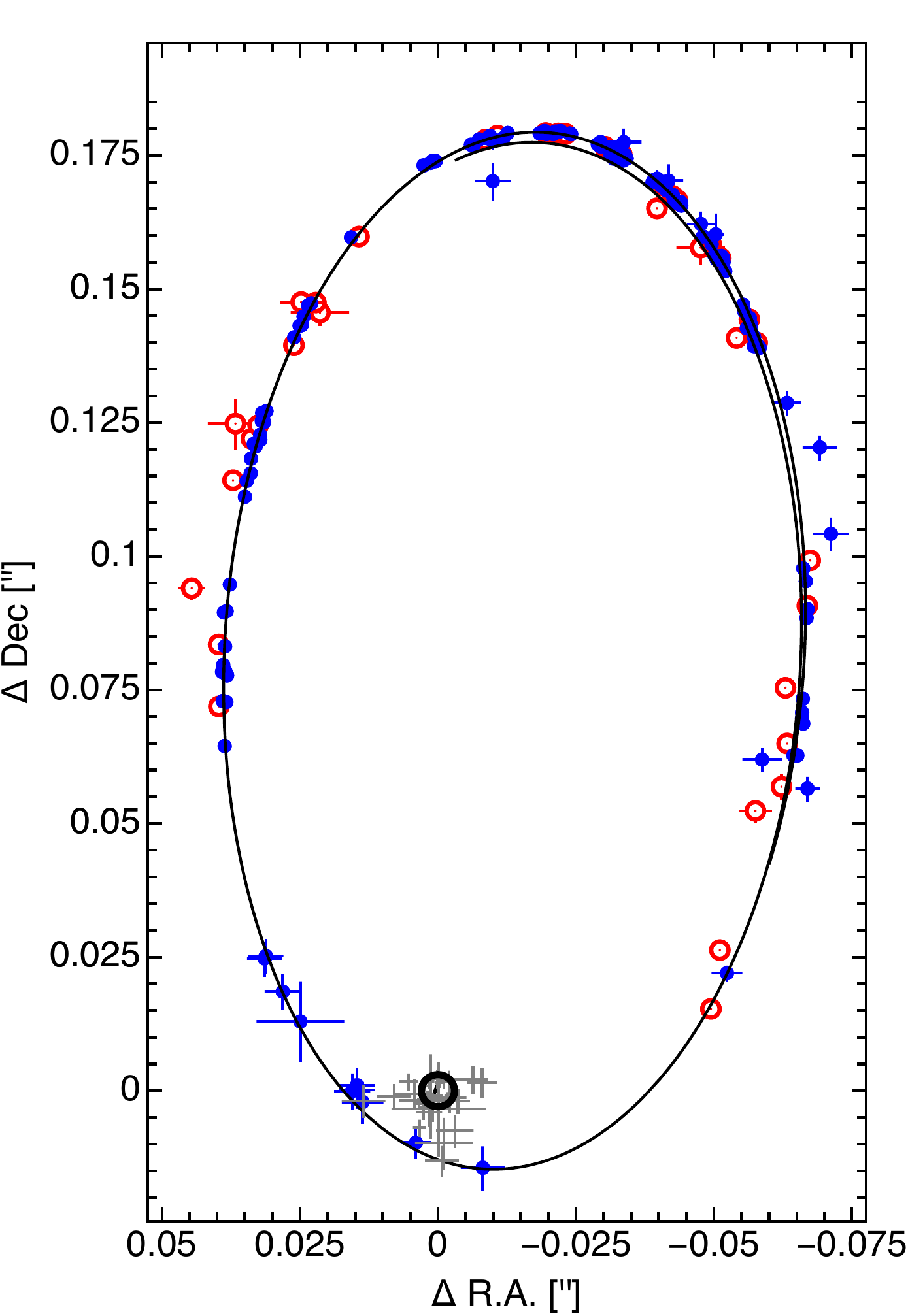}
\includegraphics[width=0.475\linewidth]{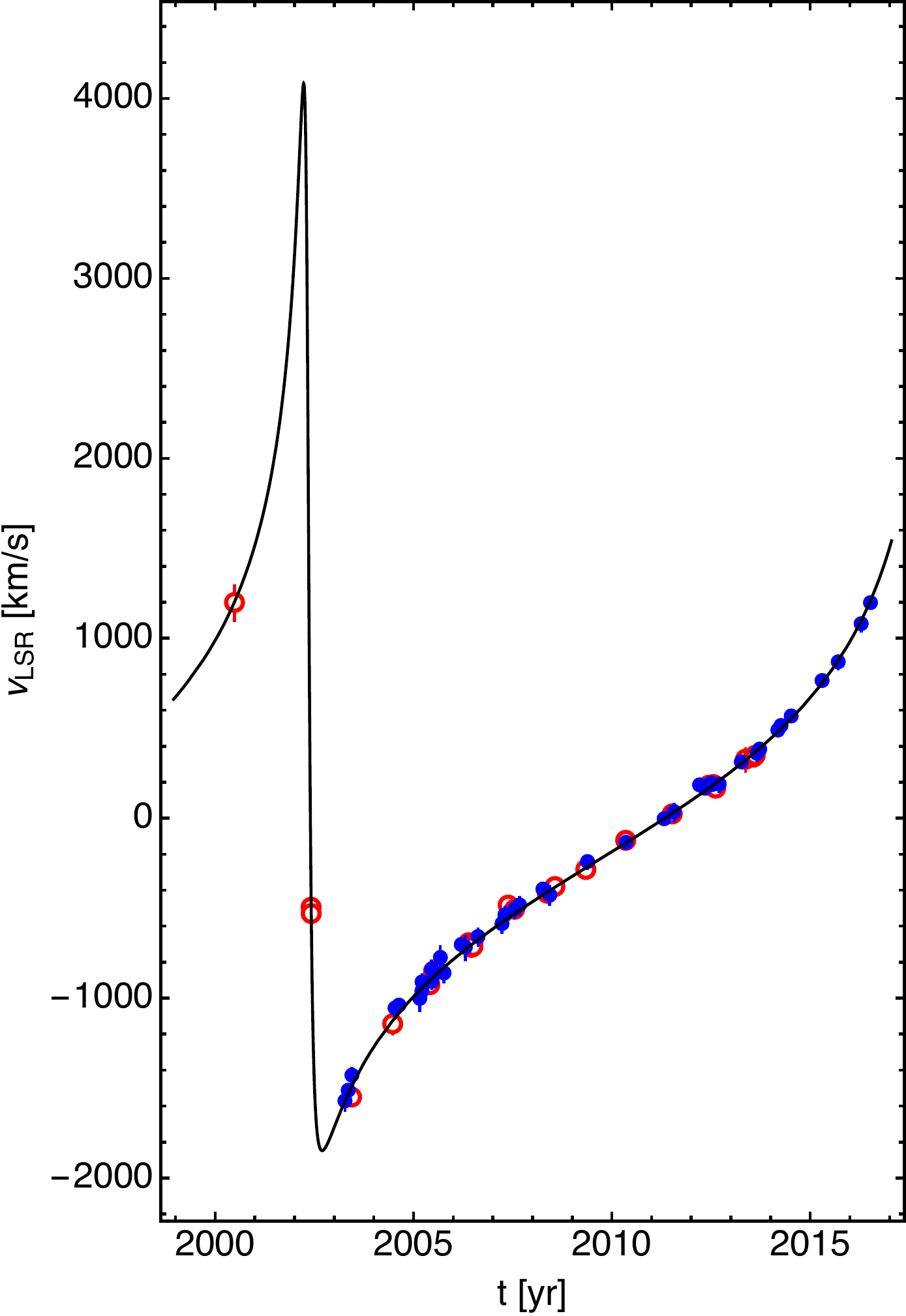}
\caption{The orbit of the star S2. Left: The measured positions plotted in the plane of sky. The blue data are from the VLT (before 2002: from the NTT), the red data are from \cite{2016ApJ...830...17B} corrected for the difference in reference coordinate system. The gray data points are positions at which flares have been recorded. The black ellipse is the best fitting orbit, the position of the mass is denoted by the black circle. Note that the fitting procedure matches the functions $\alpha(t)$ and $\delta(t)$, i.e. it does not only match the positions in the plane of sky but rather also in time. The plotted ellipse does not close, since there is a small residual drift motion of the fitted mass in the reference frame. The physical model is purely Keplerian. Right: The measured radial velocities as a function of time. The same best fitting orbit as in the left panel is denoted by the black line.}
\label{fig_fit_s2}
\end{figure*}

We first fit the orbit of the star S2 without using any coordinate system prior information (row 1 in table~\ref{tab_fit_s2}). The 2D coordinate system parameters from the fit are consistent with what we expect from the priors, and we can repeat the fit using the priors as additional constraints (row 2). We also repeat the fit from \cite{2009ApJ...707L.114G} in row 3 of table~\ref{tab_fit_s2}, which includes into our data set the publicly available S2 data from \cite{2016ApJ...830...17B}. This comes at the cost of having to solve for four additional parameters, namely the difference between the two reference frames. One can see from the parameter errors that the additional information by including the Keck data is countered by the inclusion of four additional free parameters. The combination of the two data sets does therefore not constrain the potential further in a substantial way, although one might have expected that the Keck data would help for the years 1995 - 2001, where our data set is based on lower Strehl ratio Speckle-imaging at the smaller 3.6m ESO NTT on La Silla. Figure~\ref{fig_fit_s2} shows the positional and velocity data and the best-fitting orbit, for the case where the priors and the Keck data have been used in addition to our raw data set. We refer to it as the "combined" fit. It constitutes our best estimate for the S2 orbit and the corresponding potential:
\begin{eqnarray}
M &=& 4.35 \pm 0.13 \times 10^6 \, M_\odot \nonumber \\ 
R_0 &=& 8.33 \pm 0.12 \, \mathrm{kpc}
\end{eqnarray}
These errors are only the formal fit errors, the additional systematic errors for S2 are determined in section~\ref{syserrS2}. As a cross check, we also fit the combined S2 data set without using any priors (row 4 in table~\ref{tab_fit_s2}). Again, the resulting parameters do not deviate significantly between a fit using the priors (row~3) or not (row~4). We conclude that the inclusion of the Keck data does not introduce any significant systematic errors related to the different coordinate system.

\begin{figure*}
\centering
\includegraphics[width=0.32\linewidth]{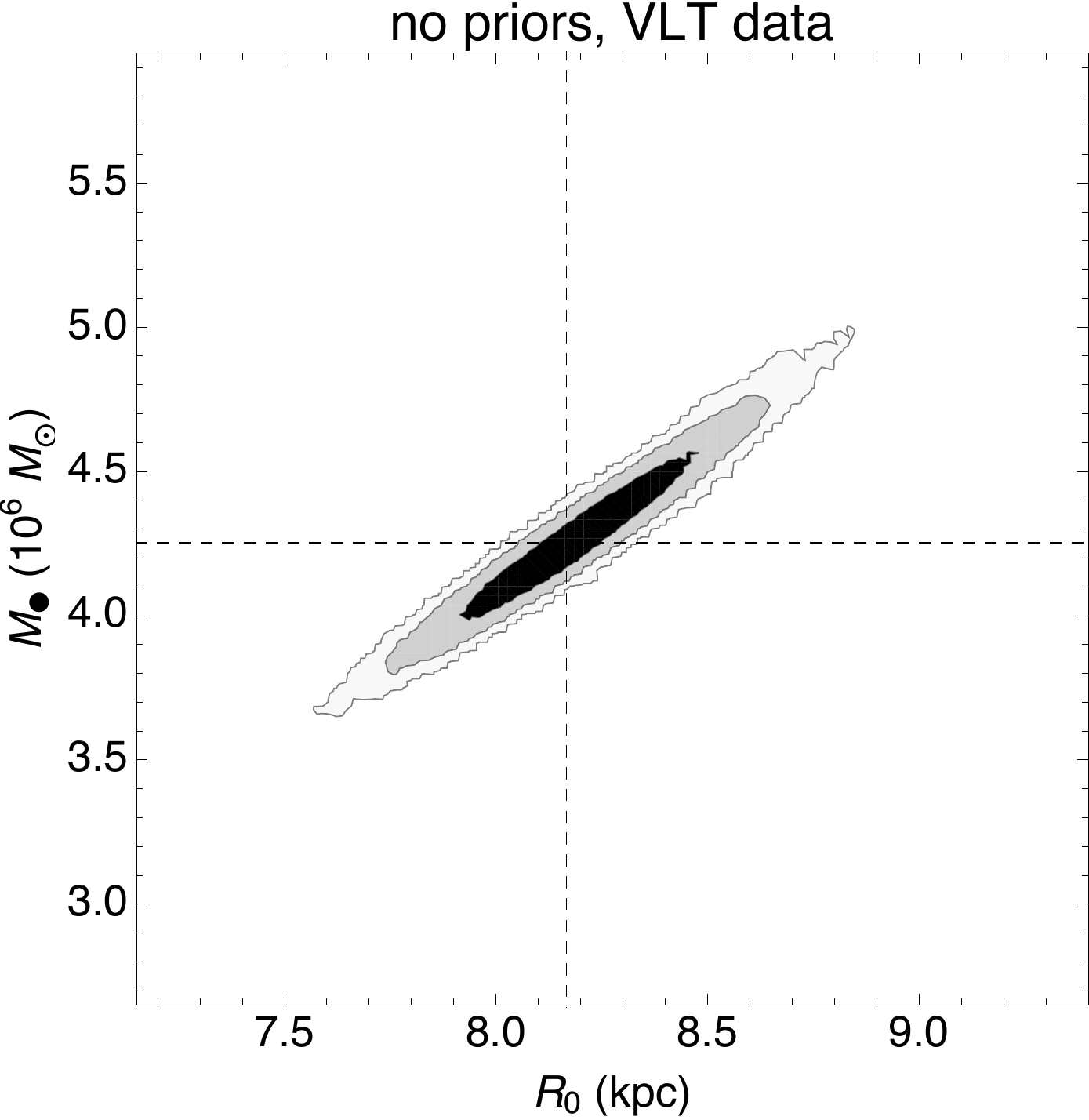}
\includegraphics[width=0.32\linewidth]{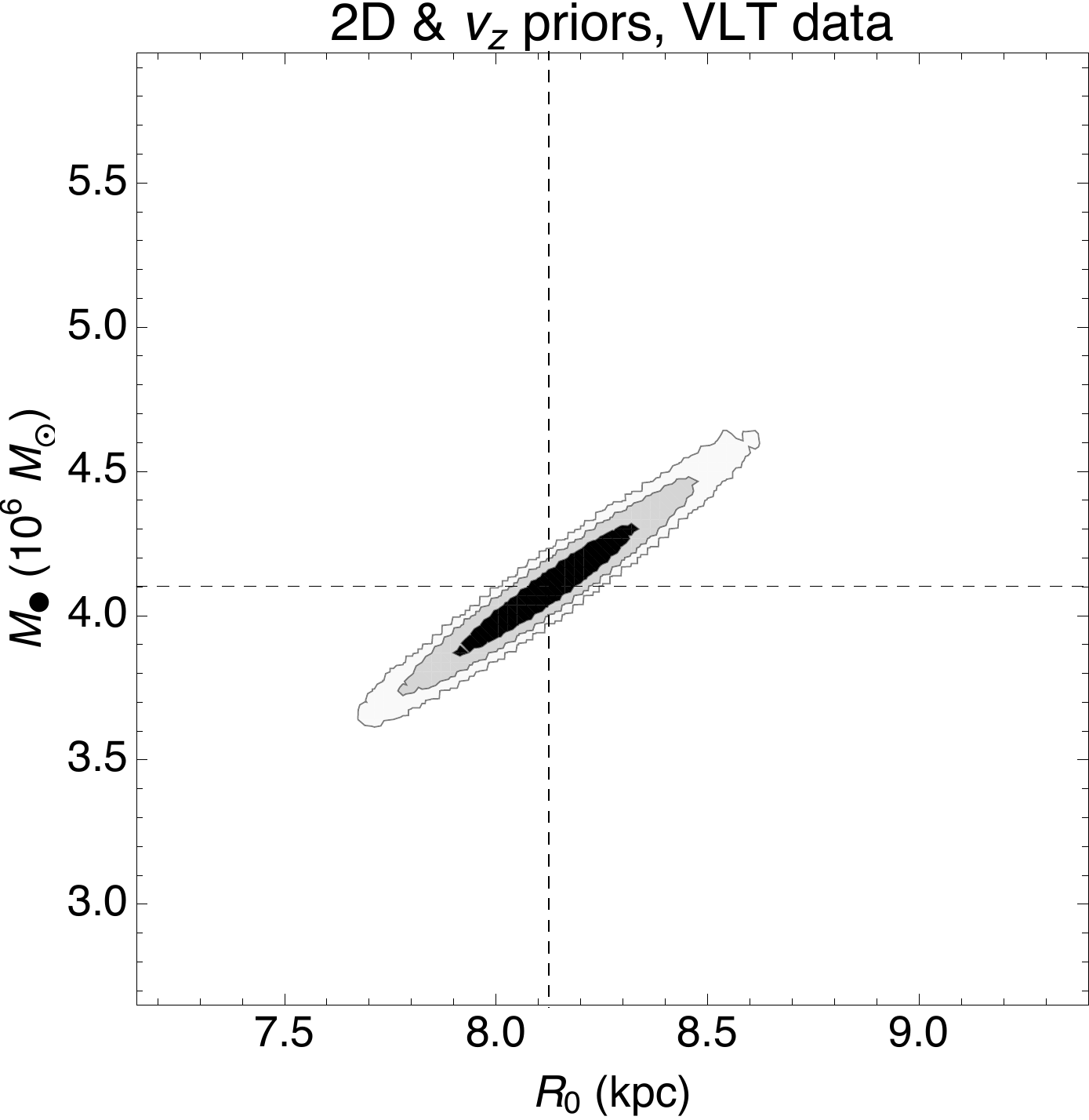}
\includegraphics[width=0.32\linewidth]{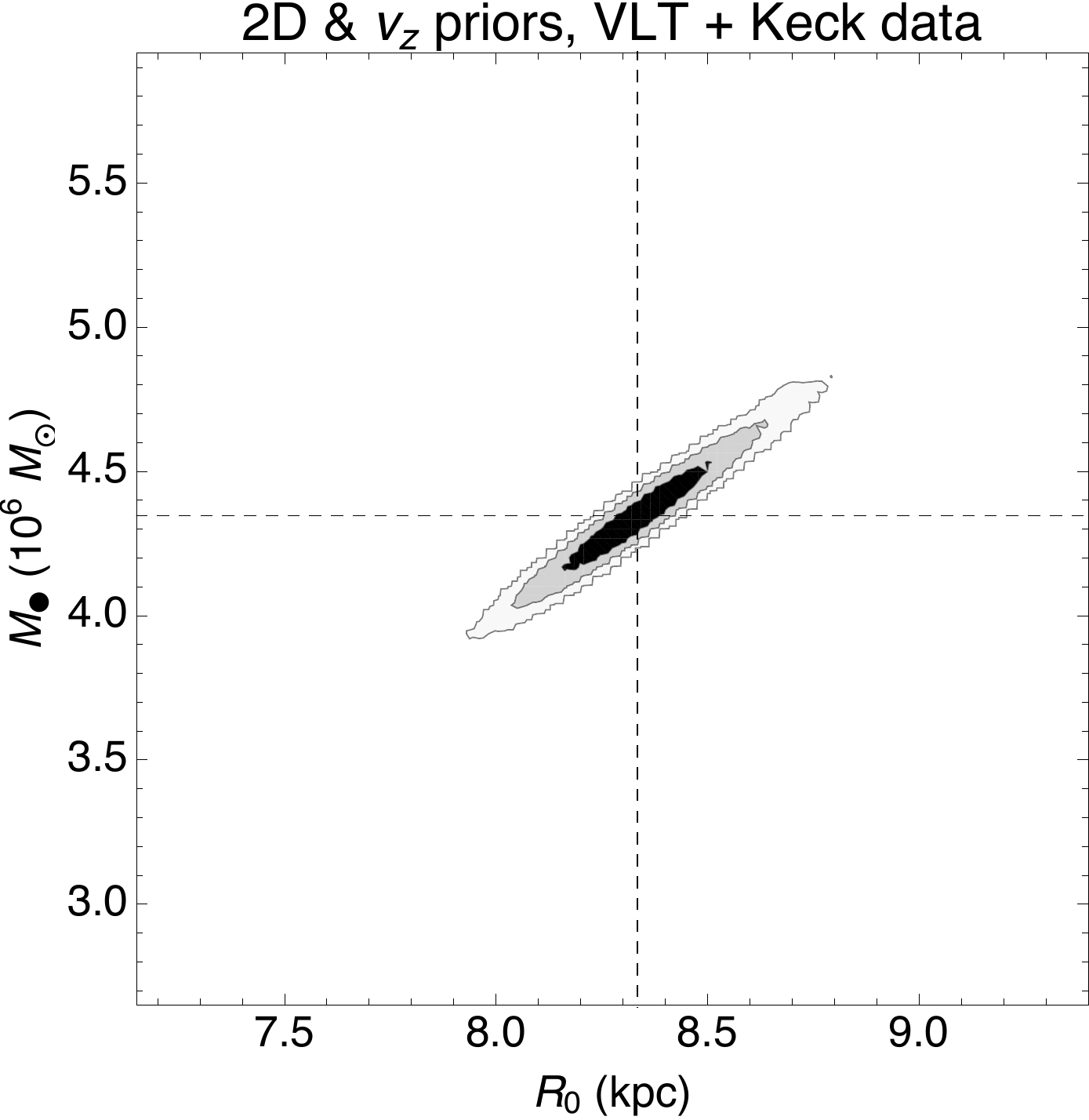}
\caption{Mass of and distance to Sgr~A* from the orbit of S2. The three panels show projections of the respective Markov chains into the mass-distance plane, giving contours at the 1-, 2- and 3-$\sigma$ level. The dashed lines mark the best fit values. The left panel is for a fit without prior information (row 1 in table~\ref{tab_fit_s2}), the middle panel includes the priors and thus leads to smaller parameter uncertainties (row 2 in table~\ref{tab_fit_s2}). The right panel in addition uses the Keck data from \cite{2008ApJ...689.1044G}, which leads to a small shift of the best fitting parameters with virtually unchanged uncertainties (row 3 in table~\ref{tab_fit_s2}).}
\label{fig_massdist_s2}
\end{figure*}

After using a classical minimization routine, we also run a Markov chain Monte Carlos (MCMC), which we started at the previously determined best fit position and which we ran for at least $2 \times 10^5$ steps. Since the posterior distribution is compact, this approach is sufficient. As expected, the chain never hit any point in parameter space with a smaller $\chi^2$. The posterior distribution has more information about the parameter uncertainties.  Table~\ref{tab_fit_s2} also gives the associated errors. The parameter uncertainties obtained from the formal error matrix and from the MCMC agree. In figure~\ref{fig_massdist_s2} we show the marginal posterior distribution of $R_0$ and $M_\mathrm{MBH}$ for the fits given in rows 1 - 3 of table~\ref{tab_fit_s2}. A more complete view of the chain output is given in the appendix in figure~\ref{fig_full_mcmc}, where we show the two-dimensional projections for the 13 parameters (excluding the four needed to describe the coordinate system mismatch between our data and the Keck data). The figure shows explicitly that the posterior distribution is compact, and that all parameters are well-constrained.

The mass of and distance to Sgr~A* are highly correlated parameters. Using the S2 data, for a given distance, the corresponding mass is
\begin{equation}
M(R) = (4.005 \pm 0.033) \times 10^6 M_\odot \times (R_0/8\,\mathrm{kpc})^{2.00} \,\, .
\end{equation}
The mass uncertainty for a given distance is below 1\%.

\begin{table*}
\caption{The gravitational potential based on orbital fitting. The first four fits use the S2 data and differ in whether or not the Keck data are used, and whether or not we include coordinate system priors in the fit. For each fit we report the best fitting parameters and the associated uncertainties as obtained by the error matrix. For S2, we also report the uncertainties as obtained by running a Markov chain Monte Carlo. The $1\sigma$ error intervals are constructed as symmetric confidence intervals around the best fitting value. Rows 5 to 7 give the same parameters as obtained from fitting S1, S9 and S13 individually. For these stars, the errors have been scaled up by the square root of the reduced $\chi^2$ (last column), corresponding to an rescaling such that reduced $\chi^2$ equals one. Row 8 presents a relativistic fit for the combined S2 data. Row 9 (bold) gives the multi-star fit using 17 stars simultaneously. The errors are taken from the Markov chain, and the reduced $\chi^2$ is smaller than 1 since before starting the fit all stars have been individually rescaled such that their respective reduced $\chi^2$ equals 1. This is our best fit overall. Row 10 is the result of the multi-star fit excluding S2.
\label{tab_fit_s2}}
{\scriptsize
\begin{center}
\begin{tabular}{l|ccc|rrrrrrrr}
\#&data & priors & type & \multicolumn{1}{c}{$R_0$}& \multicolumn{1}{c}{$M_\mathrm{MBH} $}&
 \multicolumn{1}{c}{$\alpha $}& \multicolumn{1}{c}{$\delta $}& \multicolumn{1}{c}{$v_\alpha $}&
 \multicolumn{1}{c}{$v_\delta $}& \multicolumn{1}{c}{$v_z$ } & \multicolumn{1}{c}{r. $\chi^2$}\\
&&&&  \multicolumn{1}{c}{(kpc)} & \multicolumn{1}{c}{$(10^6 M_\odot)$} &  \multicolumn{1}{c}{(mas)} &  \multicolumn{1}{c}{(mas)} &  \multicolumn{1}{c}{($\mu$as/yr)} &  \multicolumn{1}{c}{($\mu$as/yr)} &  \multicolumn{1}{c}{(km/s)}&\\
\hline
\hline
1&S2, VLT& none& Kepl. 				& $8.17\pm0.20$& $4.25\pm0.20$&$0.23\pm0.39$&$-2.10\pm0.61$&$88\pm40$&$-2\pm63$&$28.3\pm7.0$&1.19\\
&\multicolumn{3}{|r|}{MCMC errors} & $\substack{+0.17\\-0.23}$ &$\substack{+0.18\\-0.22}$&$\substack{+0.39\\-0.40}$&$\substack{+0.53\\-0.69}$&$\substack{+41\\-40}$&$\substack{+67\\-61}$&$\substack{+6.4\\-8.2}$& \TopStrut \BottomStrut \\
\hline 
2&S2, VLT& 2D, $v_z$ & Kepl. 		& $8.13\pm0.15$& $4.10\pm0.16$&$-0.31\pm0.34$&$-1.29\pm0.44$&$78\pm37$&$126\pm47$&$8.9\pm4.0$&1.28\\
&\multicolumn{3}{|r|}{MCMC errors} & $\substack{+0.13\\-0.16}$&$\substack{+0.14\\-0.16}$&$\substack{+0.34\\-0.34}$&$\substack{+0.43\\-0.45}$&$\substack{+37\\-37}$&$\substack{+47\\-46}$&$\substack{+3.9\\-4.0}$&  \TopStrut \BottomStrut \\
\hline 
3&S2, comb.& 2D, $v_z$ & Kepl. 	&$8.33\pm0.12$& $4.35\pm0.13$&$0.96\pm0.21$&$-1.28\pm0.32$&$-45\pm23$&$120\pm33$&$5.0\pm3.6$&1.48\\
&\multicolumn{3}{|r|}{MCMC errors} & $\substack{+0.12\\-0.12}$& $\substack{+0.13\\-0.14}$&$\substack{+0.21\\-0.21}$&$\substack{+0.32\\-0.34}$&$\substack{+34\\-33}$&$\substack{+23\\-23}$&$\substack{+3.5\\-3.7}$&  \TopStrut \BottomStrut \\
&$\Delta$ sys.&&&&&$-0.83\pm0.21$&$-0.81\pm0.21$&$462\pm21$&$-85\pm20$&& \\
&\multicolumn{3}{|r|}{MCMC errors} &&& $\substack{+0.22\\-0.20}$&$\substack{+0.22\\-0.22}$&$\substack{+19\\-21}$&$\substack{+21\\-22}$  \TopStrut \BottomStrut  \\
\hline 
4&S2, comb. & none & Kepl. 	&$8.17\pm0.15$& $4.30\pm0.15$&$1.49\pm0.24$&$-2.41\pm0.49$&$-34\pm24$&$24\pm44$&$11.5\pm5.4$&1.41\\
&\multicolumn{3}{|r|}{MCMC errors} & $\substack{+0.15\\-0.15}$& $\substack{+0.16\\-0.15}$&$\substack{+0.23\\-0.25}$&$\substack{+0.49\\-0.49}$&$\substack{+25\\-24}$&$\substack{+44\\-45}$&$\substack{+5.5\\-5.5}$&  \TopStrut \BottomStrut \\
&$\Delta$ sys.&&&&&$-1.06\pm0.23$&$-0.50\pm0.22$&$485\pm22$&$-114\pm21$&& \\
&\multicolumn{3}{|r|}{MCMC errors} &&& $\substack{+0.20\\-0.23}$&$\substack{+0.23\\-0.23}$&$\substack{+22\\-20}$&$\substack{+22\\-22}$  \TopStrut \BottomStrut  \\
\hline 
\hline 
5&S1& 2D, $v_z$ & Kepl. 			& $8.47 \pm 0.18$ & $4.45 \pm 0.28$ & $-0.89 \pm 1.27$ & $-0.19 \pm 1.31$ & $80 \pm 139$ & $17 \pm 143$ & $-0.1 \pm 7.4$&  2.21 \\
\hline
6&S9& 2D, $v_z$ & Kepl. 			& $8.08 \pm 0.78$ & $4.04 \pm 1.26$ & $0.21 \pm 1.51$ & $0.10 \pm 1.52$ & $-38 \pm 164$ & $-17 \pm 165$ & $0.0 \pm 8.3$&  2.73 \\
\hline
7&S13& 2D, $v_z$ & Kepl. 			& $8.74 \pm 0.97$ & $4.84 \pm 1.59$ & $-0.22 \pm 2.60$ & $2.22 \pm 2.57$ & $86 \pm 291$ & $-296 \pm 277$ & $-3.1 \pm 15.5$&  10.6 \\
\hline
\hline 
8&S2, comb.& 2D, $v_z$ & GR 			& $8.41 \pm 0.13$ & $4.43 \pm 0.14$ & $0.58 \pm 0.21$ & $-1.31 \pm 0.33$ & $-16 \pm 23$ & $120 \pm 34$ & $1.5 \pm 3.6$&  1.47 \\
\hline
\hline
\bf 9& \bf Multi&\bf  {2D,} ${\mathbf v_z}$ &\bf Kepl. 		&$\mathbf{8.32 \pm 0.07}$ &$\mathbf{4.28 \pm 0.10}$ &$\mathbf{-0.08 \pm 0.37}$ &$\mathbf{-0.89 \pm 0.31}$ &$\mathbf{39 \pm 41}$ &$\mathbf{58 \pm 37}$ &$\mathbf{14.2 \pm 3.6}$&\bf  0.98\\
10&w/o S2& 2D, $v_z$ & Kepl. 		& $8.19 \substack{+0.16\\-0.11}$ & $4.08 \substack{+0.25\\-0.14}$ & $0.55 \substack{+0.65\\-0.62}$ & $-0.26 \substack{+0.64\\-0.60}$ & $-83 \substack{+0.69\\-0.73}$ & $-22 \substack{+0.65\\-0.70}$ & $7.0 \substack{3.7\\-3.6}$& 0.97\BottomStrut\\
\hline
\hline
\end{tabular}
\end{center}
}
\end{table*}

The fit of the MPE-only data set without priors in row 1 of table~\ref{tab_fit_s2} yields a somewhat large radial velocity of the central mass of $28\,$km/s. This might be connected to a systematic error of measuring radial velocities, see section~\ref{syserrS2}. 
In order to decouple the fit from such a bias, but still being able to profit from the 2D priors, we have repeated the fit with 2D priors only. This fit yields again a systemic radial velocity of around $25\,$km/s. The distance estimate is $R_0=8.35\,$kpc then, very close to what our fiducial fit yields (row 3).

We also tested whether allowing for a constant rotation speed of the coordinate system would make a difference in the fits. Using the combined S2 data set, the parameter corresponding to rotation has its best fit value at $(-0.006 \pm 0.017)^\circ$/yr, not significantly deviating from zero. The distance estimate is basically unchanged \mbox{$R_0=8.34\,$kpc}. We conclude that we can neglect rotation. Furthermore, from the definition of the coordinate system we can set a prior on the coordinate system rotation of $0.00004^\circ$/yr \citep{2015MNRAS.453.3234P}, which is essentially the same as fixing it at zero. For simplicity we do the latter.

\subsubsection{Relativistic fit}
\label{relfit}

Using a general relativistic orbit model, the S2 based mass and distance are
\begin{eqnarray}
M &=& 4.43 \pm 0.14 \times 10^6 \, M_\odot \nonumber \\ 
R_0 &=& 8.41 \pm 0.13 \, \mathrm{kpc} \,\,.
\end{eqnarray}

We note that the values increase moderately compared to the respective Keplerian fits. It was already noted by \cite{2006ApJ...639L..21Z} that the Keplerian fit yields biased parameter estimates for a relativistic orbit, although the size (and sign) of these biases has not yet been studied systematically.

In the orbit fitting tool, we can check the effect of the four relativistic corrections implemented independently. After rescaling the error bars such that the Keplerian (combined) S2 fit yields a reduced $\chi^2$ of 1, we tested models with the different effects, or combination of those, one by one. This yielded reduced $\chi^2$ values between $\approx 0.987$ and $\approx 1.020$. For the 414 degrees of freedom a $1$-$\sigma$-significant deviation would be reached for $\Delta \chi^2_\mathrm{red} = \sqrt{2/\mathrm{d.o.f.}}=0.070$ \citep{2010arXiv1012.3754A}. Hence, we cannot distinguish between any of these models, and cannot detect any of the leading order relativistic effects: Schwarzschild precession, gravitational redshift, relativistic Doppler effect and Roemer delay. The Keplerian description continues to suffice.

\subsubsection{Limits on an extended mass component}
\label{extfit}

Fitting the combined S2 data using an additional, extended mass component with a Plummer profile
\begin{equation}
 \rho(r) = \frac{3}{8 \pi} \, M_\mathrm{ext} \, r_s^{-3} \left( \frac{r}{r_s} \right)^{-5/2}  
\end{equation}
with a scale radius of $r_s = 0.4"$ yields that $-0.4 \pm 1.2 \%$ of the mass of the MBH is in the extended component, where we have allowed $M_\mathrm{ext}$ to also take negative values.
This corresponds to $-0.3 \pm 0.7 \%$ between pericenter and apocenter of the S2 orbit, the radial range where our data are sensitive to
an additional mass component. Changing $r_s$ to $0.125"$ yields a very similar result, with  $-0.5 \pm 0.8 \%$  being in the extended component or  $-0.3 \pm 0.5 \%$ inside the S2 orbit.
We also used a power law density profile with $\rho(r) \propto r^{-7/4}$, making the extended mass component inside the S2 orbit   $-0.5 \pm 0.8 \%$.

We conclude that our data are consistent with a pure point mass, and can place a conservative upper limit on a possible extended component inside the S2 orbit at $1\%$ of the mass of the MBH.

\subsubsection{Systematic errors for S2}
\label{syserrS2}

\begin{figure*}
\centering
\includegraphics[width=0.23\linewidth]{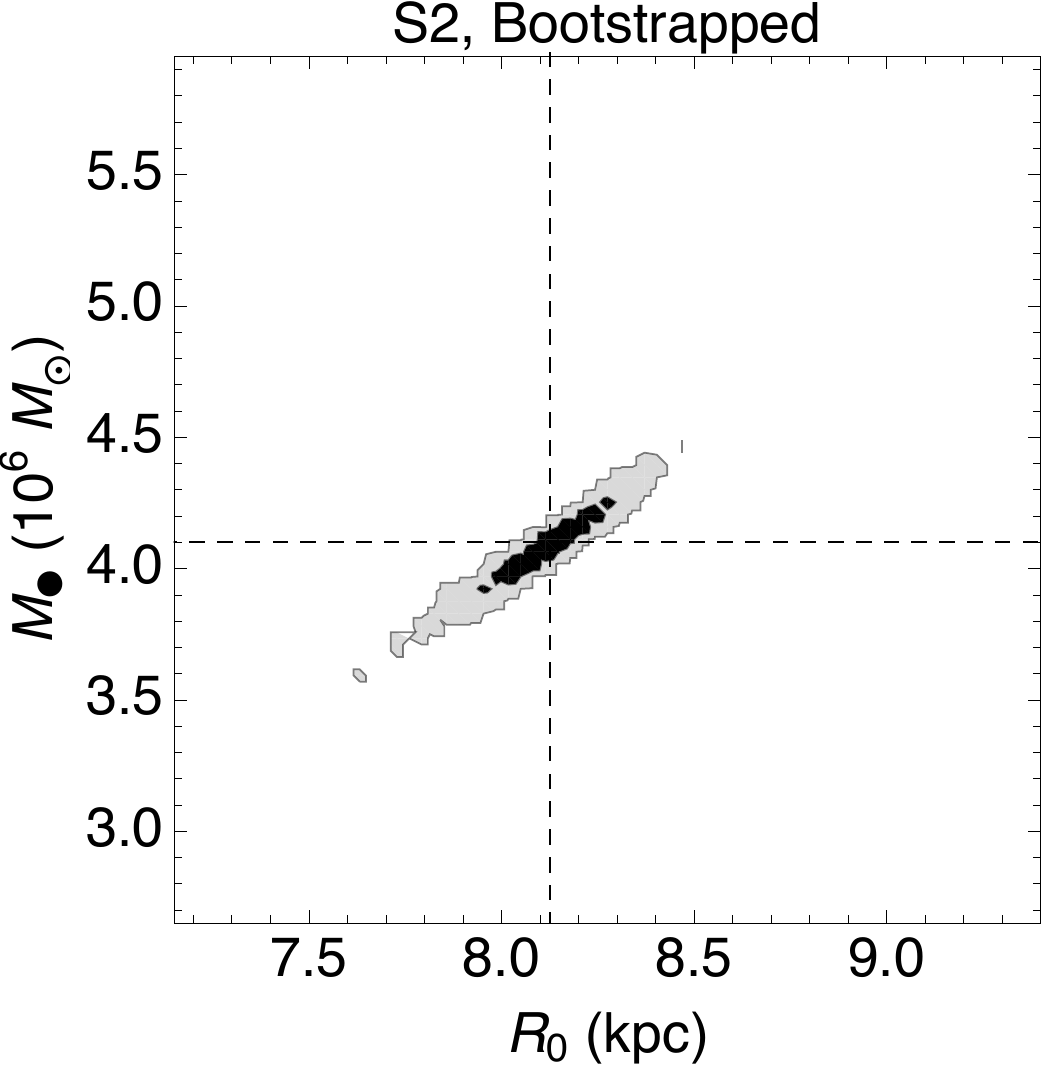}
\includegraphics[width=0.23\linewidth]{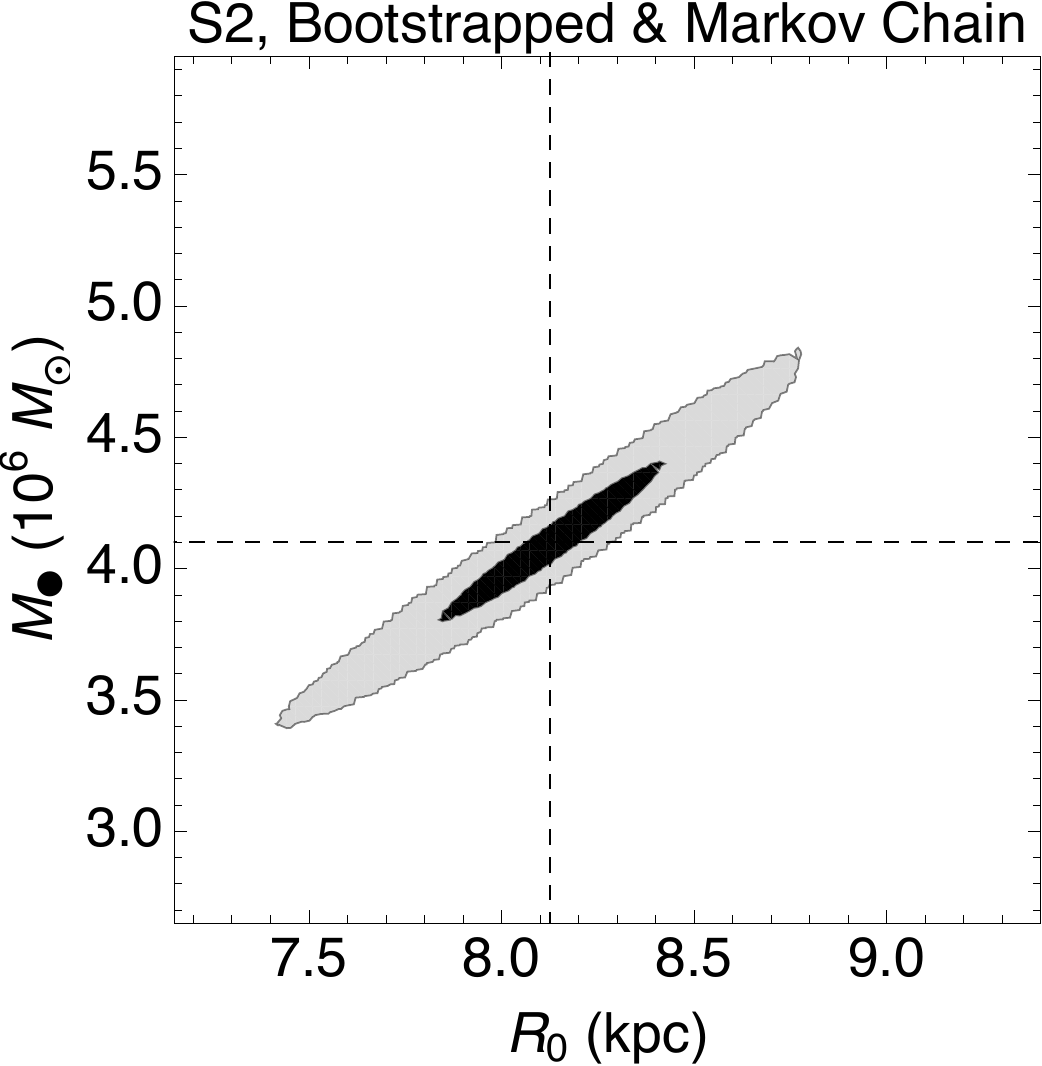}
\includegraphics[width=0.23\linewidth]{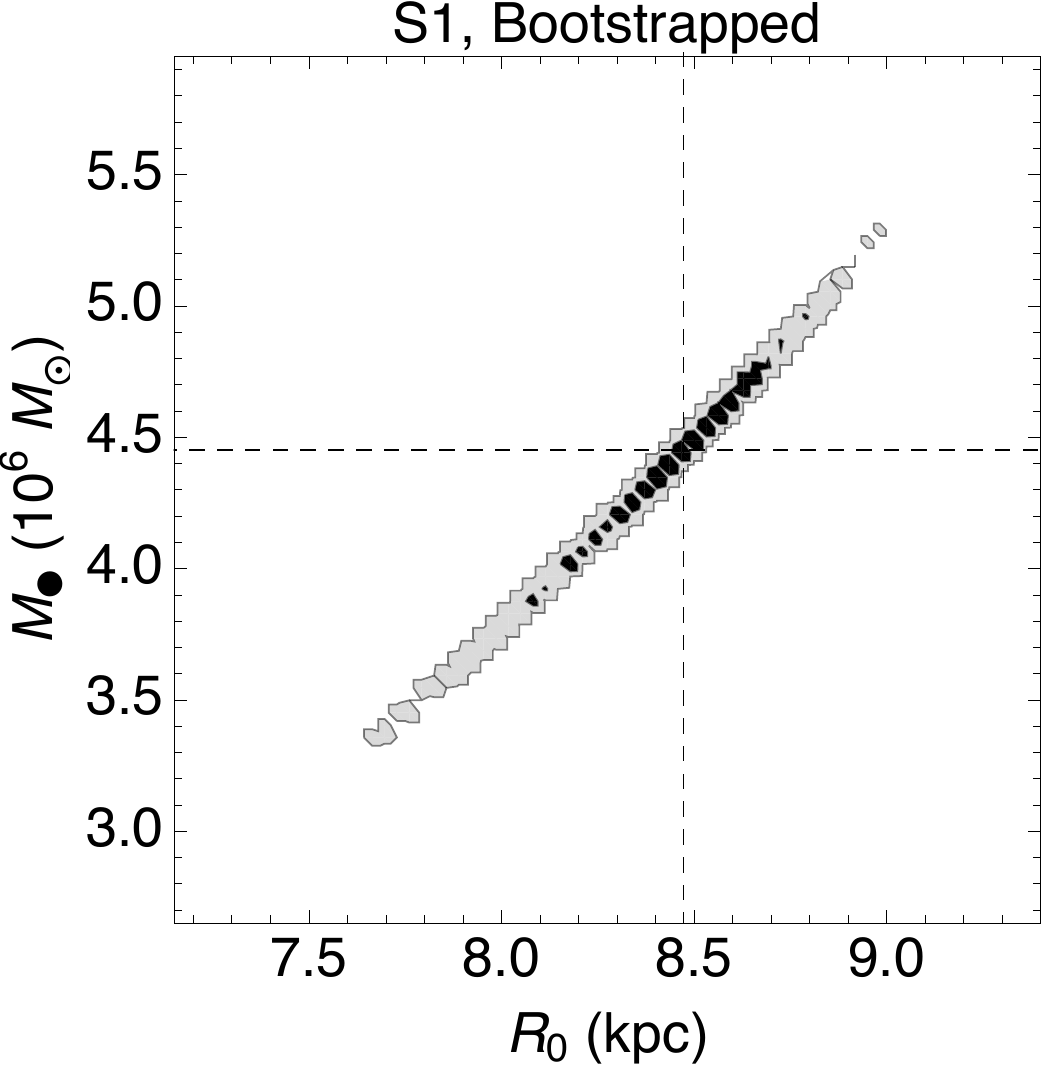}
\includegraphics[width=0.23\linewidth]{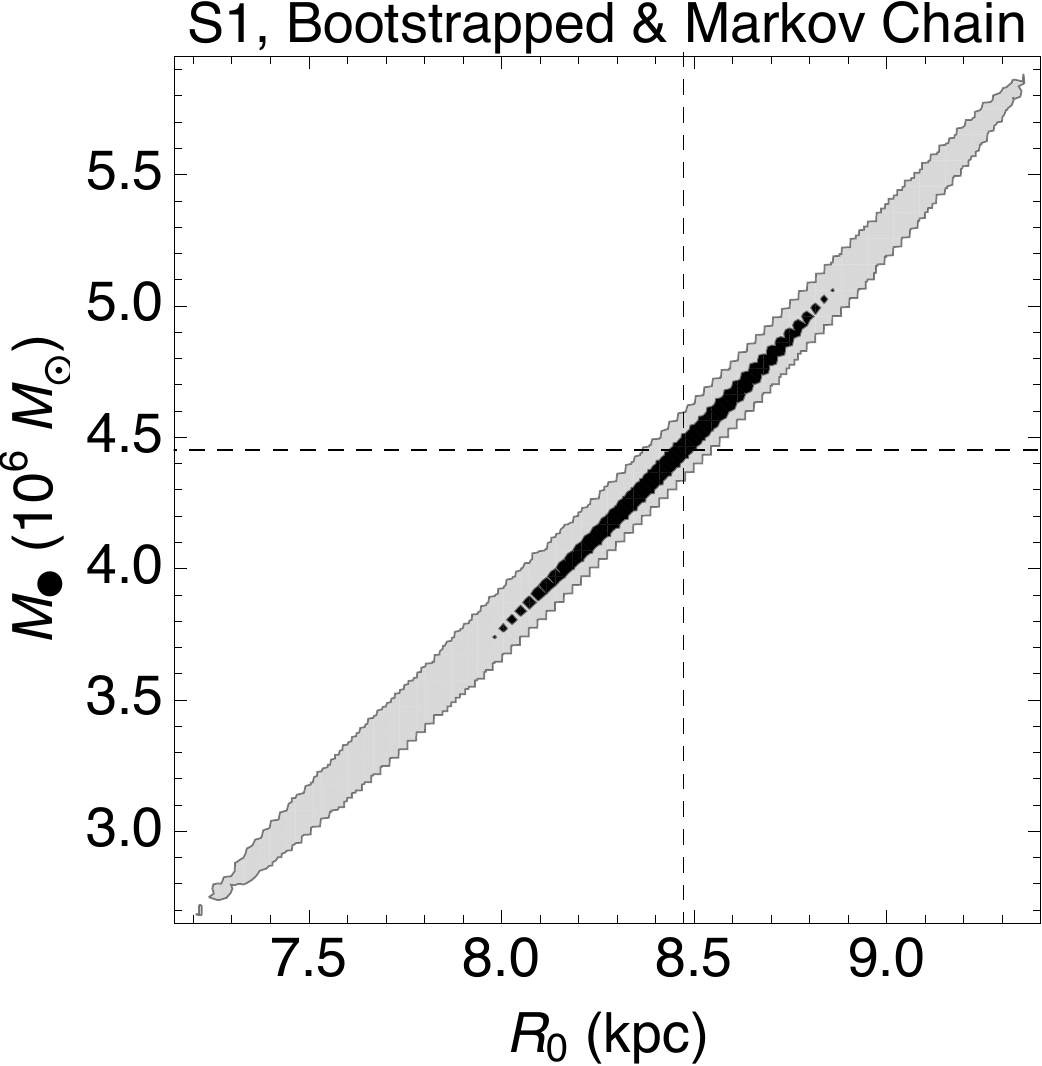}
\caption{Parameter uncertainties from bootstrapping shown for $M$ and $R_0$. Left panels: For S2. Right panels: For S1. For both stars the left panel shows the pure uncertainties from the 1000 bootstrap samples, and the right one the combined taking into account in addition the statistical errors from the Markov chain. The dashed lines mark the best fit values.}
\label{fig_bootstrap}
\end{figure*}

A main source of uncertainty in \cite{2009ApJ...692.1075G} was the weight of the S2 data in 2002. A fit using S2 only leaving out the 2002 data yielded a distance value as low as $R_0 = 7.4\,$kpc, while using the 2002 data with their full weights yielded $R_0=8.9\,$kpc. The range of values of $R_0$ for the updated (combined VLT and Keck) data set now is $8.24\,\mathrm{kpc} < R_0 < 8.84\,\mathrm{kpc}$, i.e. it has reduced by more than a factor 2. 

The influence of individual data points on the fit result can be checked by bootstrapping. For that we created 1000 bootstrapped files by
 drawing randomly with replacement as many data points from a given star's data as there are measurements. Some data points are thus repeated in the bootstrapped file, others are omitted. These mock data sets are then fit in the standard way, and the distribution of best fitting parameters is a measure for the uncertainty in the data. Fig.~\ref{fig_bootstrap} shows in the left most panel the results for the (VLT-only) S2 data set. The associated error bars are $\pm 0.13\,$kpc
for $R_0$ and $\pm 0.13\times10^6 M_\odot$ for $M$, i.e. comparable to the statistical fit uncertainties. The second panel in fig.~\ref{fig_bootstrap} shows the combined uncertainties from bootstrapping and the Markov chain. For that figure we assumed that the statistical fit errors at the best fit position are valid at each point of the bootstrap.

While we expect instrumental systematics in $v_\mathrm{LSR}$ to average out, the shape of the stellar absorption lines for the massive B dwarfs might be affected by stellar winds. This results in a systematic difference of measured radial velocity and true radial velocity of the center of mass of the star. We estimate that such effects could bias the measurements at the $20\,$km/s level. For S2, a star for which we have measured positive and negative radial velocities, this would be absorbed into the radial motion of the coordinate system $v_z$, and indeed, the S2 fits without prior information (rows 1 and 4 in table~\ref{tab_fit_s2}) yield a value of $v_z$ of roughly that amount. 

By including the difference between the S2 fits with or without coordinate system prior information we cover thus not only the coordinate system uncertainty, but also the possible biases due to the line shape of the absorption lines. We use the mean of the half difference between the fits in rows 1 and 2, and rows 3 and 4 as contributions to the systematic error: This adds $0.05\,$kpc to the error budget.  

The difference in $R_0$ between the Keplerian and the relativistic model amounts to $0.09\,$kpc for S2. Since we have not explicitly detected relativistic effects, we include half of this in the systematic error. Similarly, we account for the models using an extended mass component, adding $0.01\,$kpc only.

Adding the contributions in squares, we estimate the systematic error of the S2-based distance estimate to be $0.17\,$kpc.

\subsection{The potential based on S1}
\label{sec_s1}

\begin{figure*}
\centering
\includegraphics[width=0.9\linewidth]{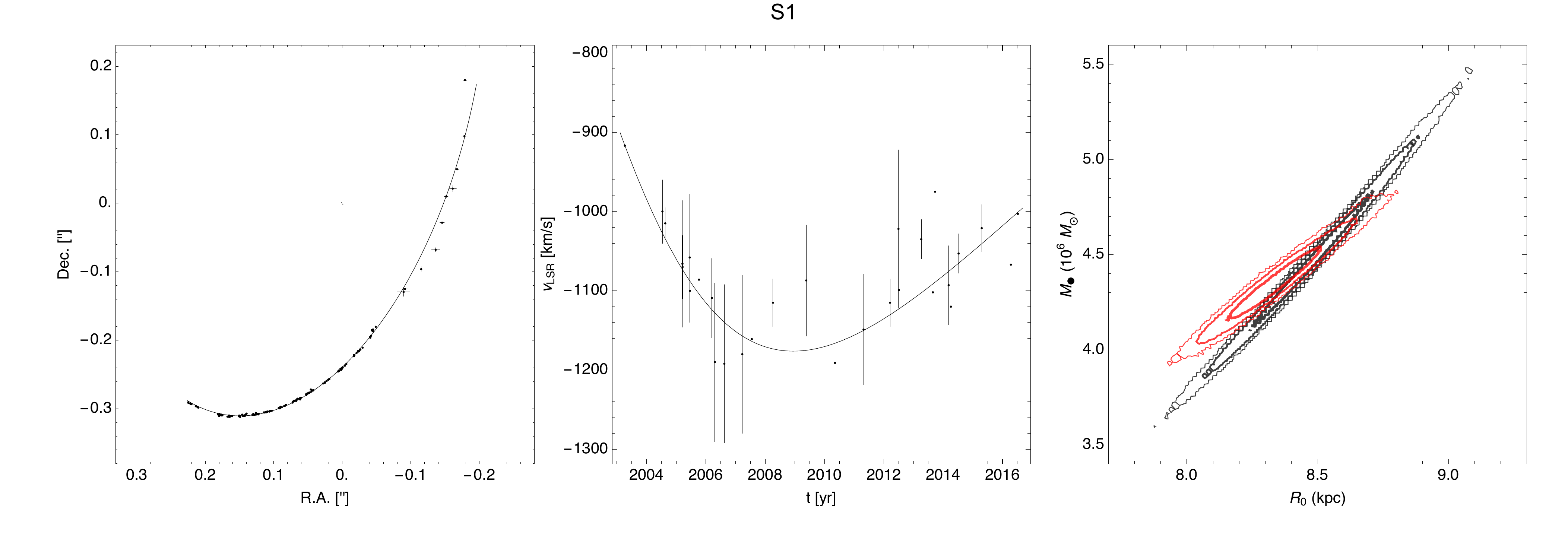}
\caption{Orbital data for S1: Left panel: The data points are the measured positions of S1 for all epochs. The line is the best fitting Keplerian orbit. The black disk marks the position of the mass as given by the orbit fit. Middle panel: Radial velocity data and the same orbit as in the left panel. Right panel: MCMC constraints on mass and distance of Sgr~A* from S1. The black contours show the constraints from S1. For comparison the S2 constraint is given by the red contours.}
\label{fig_s1}
\end{figure*}

S1 is the second-best star in terms of how much it constrains Sgr~A*'s potential (fig.~\ref{fig_s1}). We can follow the trajectory of this $m_K=14.8$ star over the full time range from 1992 to 2016, and it does not suffer from any apparent confusion. It passed the pericenter of its orbit in mid 2001. If one applies the coordinate system priors, S1 yields a similarly good constraint on $R_0$ as S2.
The best fit parameters and statistical errors for S1 are (row 5 in table~\ref{tab_fit_s2}):
\begin{eqnarray}
M &=& 4.45 \pm 0.28 \times 10^6 \, M_\odot \nonumber \\ 
R_0 &=& 8.47 \pm 0.18 \, \mathrm{kpc} \,\,.
\label{eq_fit_s1}
\end{eqnarray}
Without the priors, the S1 fit is not well constrained, and no useful constraints on $M$ and $R_0$ can be obtained. This is in contrast to the fit of S2, for which the prior information is not essential. We have three possible ways to include prior information for S1: First, we can apply the coordinate system priors as obtained from \cite{2015MNRAS.453.3234P}. This assumes that the radio source Sgr~A* is the counterpart to the mass. This method of including the coordinate system prior results in the numbers given in eq.~\ref{eq_fit_s1} and row 5 in  table~\ref{tab_fit_s2}. The second option is to use the results from the S2 fit without priors as coordinate system priors for S1. This assumes that the two stars orbit the same mass. This fit yields
$M = 4.55 \pm 0.29 \times 10^6 \, M_\odot$ and
$R_0 = 8.58 \pm 0.19 \, \mathrm{kpc}$.
The third option again assumes that S1 and S2 orbit the same mass: A simultaneous fit of the two data sets yields
$M=4.69 \pm 0.14 \times 10^6 \, M_\odot $ and
$R_0 = 8.63 \pm 0.10 \, \mathrm{kpc}$. The agreement at the $1$-$\sigma$ level between these numbers shows that any way of including the prior information is fine. 

Another difference between S1 and S2 is that the error ellipse of S1 is oriented more steeply, $M \propto R_0^3$.
For the following discussion it is useful to introduce 
\begin{equation}
\mu:=\frac{M}{R_0^3} = \frac{4 \pi^2}{G} \frac{\alpha^3}{T^2} \,\,,
\label{eqMu}
\end{equation}
where $G$ is the gravitational constant, $\alpha$ the angular size of the semi-major axis and $T$ the orbital period. For a data set with astrometry points only, one cannot measure mass and distance separately, instead one has a complete degeneracy $M \sim R_0^3$, or in other words, one constrains $\mu$. When measuring $\mu$, the only quantities that enter are the angular size of the semi-major axis and the orbital period, which both can be determined from astrometry only. For a star with both astrometry and spectroscopy, $M \sim R_0^\alpha$ with $\alpha \leq 3$, for example $\alpha = 2.00$ for our S2 data set. In the case that one has only one single radial velocity and an astrometric orbit, the error ellipsoid is oriented along the $R_0$- and $\mu$-axes, and its extension in the $R_0$-direction is given by the accuracy of the radial velocity point and by how much the inclination is degenerate with $R_0$. The latter degeneracy is severe for an (almost) face-on orbit with $i\approx 0$: If one changes $R_0$ by a factor $f$, one can find a good orbit fit at $M' = f^3 M$ and $i' = i/f^2$. 

The error ellipses in the $M$-$R_0$-plane for S1 are surprising in two aspects, given that the orbital phase coverage is less than $\pi$ for S1, while the coverage is more than a full revolution for S2: First, the S1 data yield as good a constraint on $R_0$ as does S2; and second the S1 error ellipses are thinner than the ones from S2. 

The first surprise can be explained in the following way: For S1, the radial velocity did not change much over the period covered with measurements. The data can therefore be approximated by having in addition to the astrometry essentially a single, but very well measured radial velocity (with an uncertainty of around $\sigma_v / \sqrt{N_\mathrm{data}} \approx 15\,$km/s). One has thus an error ellipsoid along the $\mu-$axis. Having one single radial velocity data point with a relative error of around ($15\,$km/s)~/~($1100\,$~km/s) $ = 1.4\%$ yields a distance error in the percent regime, up to a geometry factor depending of the shape and orientation of the orbit, which is not a large factor for S1's orbit. The distance estimate comes from the comparison of the radial velocity (in km/s) with the proper motion (in mas/yr). Compared with S2, the mean (absolute) radial velocity for S1 is actually higher, such that it is plausible, that S1 can yield a similarly tight constraint on $R_0$.

The second surprise is that S1 yields a very good constraint on $\mu$, the ellipse is even tighter in the {$\mu$-direction} as is the S2 ellipse in the $M/R_0^2$ direction. To eliminate the influence of the radial velocity data, we fitted only the astrometry of S1 and S2. This yields relative uncertainties on $\mu$ of 0.30\% for S1 and 0.78\% for S2. Hence, the astrometry of S1 yields a tighter constraint on $\mu$ than the astrometry of S2. Why is that?

\begin{figure}
\centering
\includegraphics[width=0.85\linewidth]{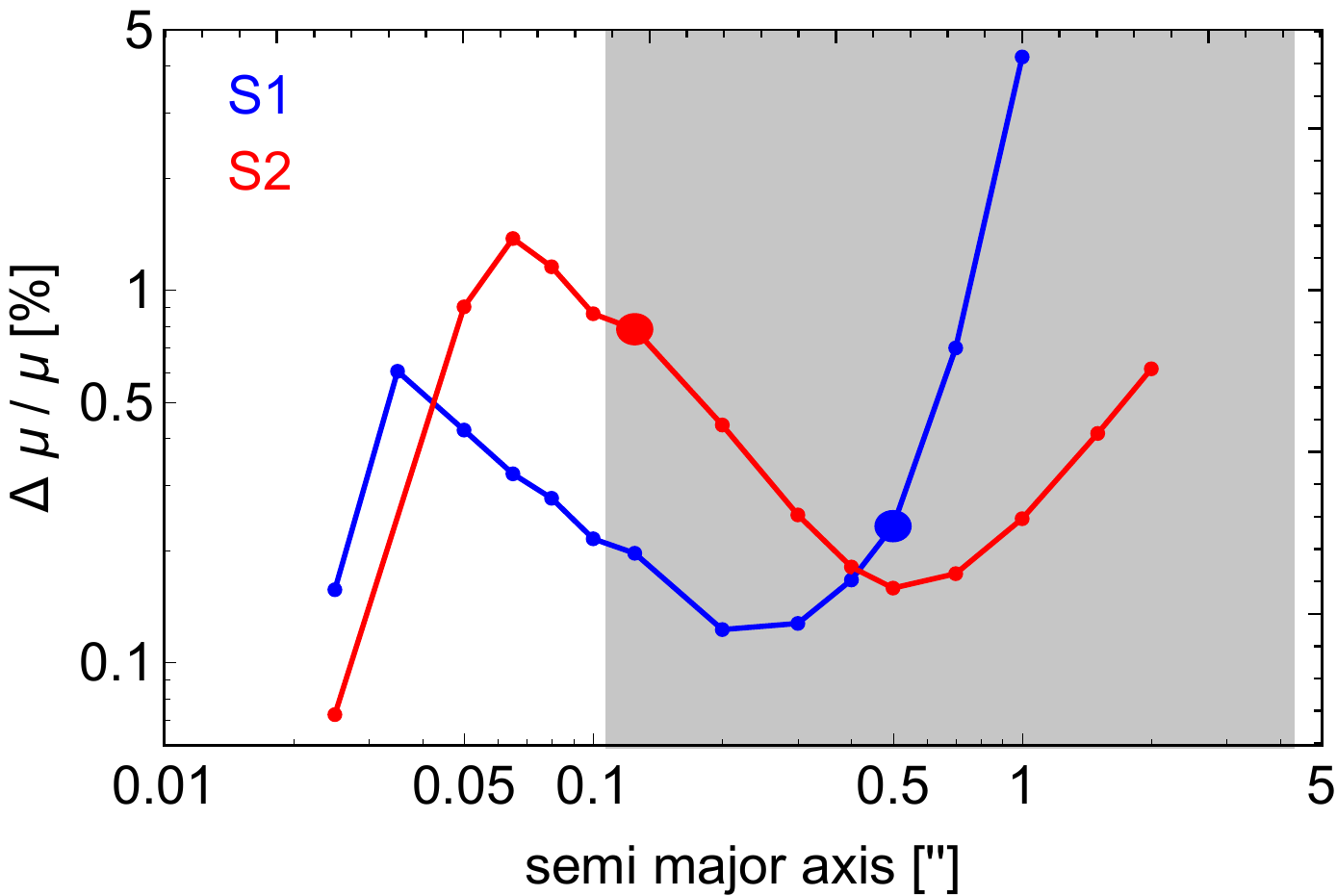}
\caption{Relative uncertainty by which $\mu=M/R_0^3$ can be constrained from a purely astrometric data set as a function of assumed semi-major axis. The shape, orientation and pericenter passage time used for simulating the orbital data were those of S1 (blue points) and S2 (red points). The thick dots mark the actual orbits of the two stars. The gray shaded area indicates the range of measured semi-major axes in table~\ref{tab_sstars_1}.}
\label{fig_mu}
\end{figure}

The answer to this question hinges on how well are angular size of the semi-major axis and the orbital period determined from an astrometric data set. To this end, we simulated a set of stellar orbits, assuming the shape, orientation and pericenter passage time of the S1 or S2 orbit, but varying the semi-major axes. For each simulated star, we created a purely astrometric data set, assuming four sampling points per year for a measurement time span of 25 years (1992 - 2016), and Gaussian position errors of $300\,\mu$as. The pericenter passages were covered in all simulated data sets. We then fit an orbit to the simulated data set and determined the relative uncertainty of $\mu$ (figure~\ref{fig_mu}). 

For both stars, the qualitative shape of the figure is the same: For very small semi-major axes, when many orbital revolutions are covered, $\Delta \mu/\mu$ is very small, and its value increases with the assumed semi-major axes. However, for semi-major axes $a \gtrsim 50\,$mas, the value of $\Delta \mu/\mu$ drops again, up to $a \approx400\,$mas, where $\Delta \mu/\mu$ reaches a minimum. For these semi-major axes the orbital phase coverage is smaller than $2\pi$. For  \mbox{$a\gtrsim 400\,$mas} the relative uncertainty $\Delta \mu/\mu$ starts increasing again, and its value increases to arbitrarily large values, since eventually one reaches the regime in which only a small arc of an orbit with a marginally significant acceleration is covered. 

The exact shape of this curve depends of course on the assumed orbit, the assumed astrometric uncertainties and the time range covered with observations. A full parameter study is beyond the scope of this work, yet the surprising behavior of $\Delta \mu / \mu$ makes our finding plausible that S1 yields a better constraint on $\mu$ than S2. S1's semi-major axes  is closer to the optimum value at which one constrains $\mu$ ideally. Note, however, that this is only true for $\mu$, and not for the mass $M$.
The mass error from S1 is somewhat larger than that from S2. 

Overall S1 and S2 yield similarly good constraints on $M$ and $R_0$ when the position of the central mass is constrained a-priori, and S1 yields even a better constraint on $\mu$ than S2. Yet, S2 is more constraining for the potential overall, since it simultaneously also constrains the position of the mass without needing prior information.

\subsubsection{Systematic errors for S1}
\label{syserrS1}
For S1, where we have essentially measured only one (constant) radial velocity, the systematic error might bias the distance estimate. Since the latter connects proper motion and radial velocity for each star in a linear fashion, a systematic error in the radial velocity will yield the same relative error in $R_0$. For S1, we get thus a $1.4\%$ systematic error, or $0.12\,$kpc. 

The different ways to include the prior information for S1 (see text after equation~\ref{eq_fit_s1}) correspond to a systematic error of $0.13\,$kpc.

Fig.~\ref{fig_bootstrap} (right panels) shows the result of bootstrapping the S1 orbit. The associated error on $R_0$ is $\substack{+0.21\\-0.36}\,$kpc. The posterior distribution is visibly asymmetric. Smaller values of $R_0$ are more likely than larger ones. The mean value of $R_0$ over the bootstrap sample is $8.41\,$kpc, $0.06\,$kpc smaller than the best fit value. This is indicative of a bias in the data. 

The quadratic sum of the systematic errors for S1 is $\substack{+0.27\\-0.40}\,$kpc.

\subsection{The potential based on multiple stars}

Not all stars for which we can determine orbits are useful for constraining the potential. For the determination of an orbit six dynamical quantities must be measured, corresponding to the number of orbital elements. In most cases, when one can determine an orbit, these are: the positions $\alpha, \delta$, the proper motions $v_\alpha, v_\delta$, the radial velocity $v_z$, and an acceleration in the plane of sky. However, also higher moments of the in-plane motion can be used instead of the radial velocity, or a change in radial velocity can be used instead of the 2D acceleration. If more than six numbers are measured, the star starts constraining the potential. For determining both mass and distance, one needs at least eight dynamical quantities.

The example of S1 shows that one can get biased estimates. Biases will mostly scale with the brightness of the stars - the fainter a star is, the more it is prone to confusion, and the more likely the orbital data are biased. But many more factors can influence how well a given star can be measured. In order to be least affected by such biases, we attempt to use the largest number of stars possible for a multi-star fit.

The next best stars after S2 and S1 in our data set are S9 and S13. We give the fit results for these stars in rows 6 and 7 of table~\ref{tab_fit_s2}. But our data set contains more stars that carry information on the potential. We identify them in the following.

\subsubsection{Selection of stars}

\begin{figure}
\centering
\includegraphics[width=0.95\linewidth]{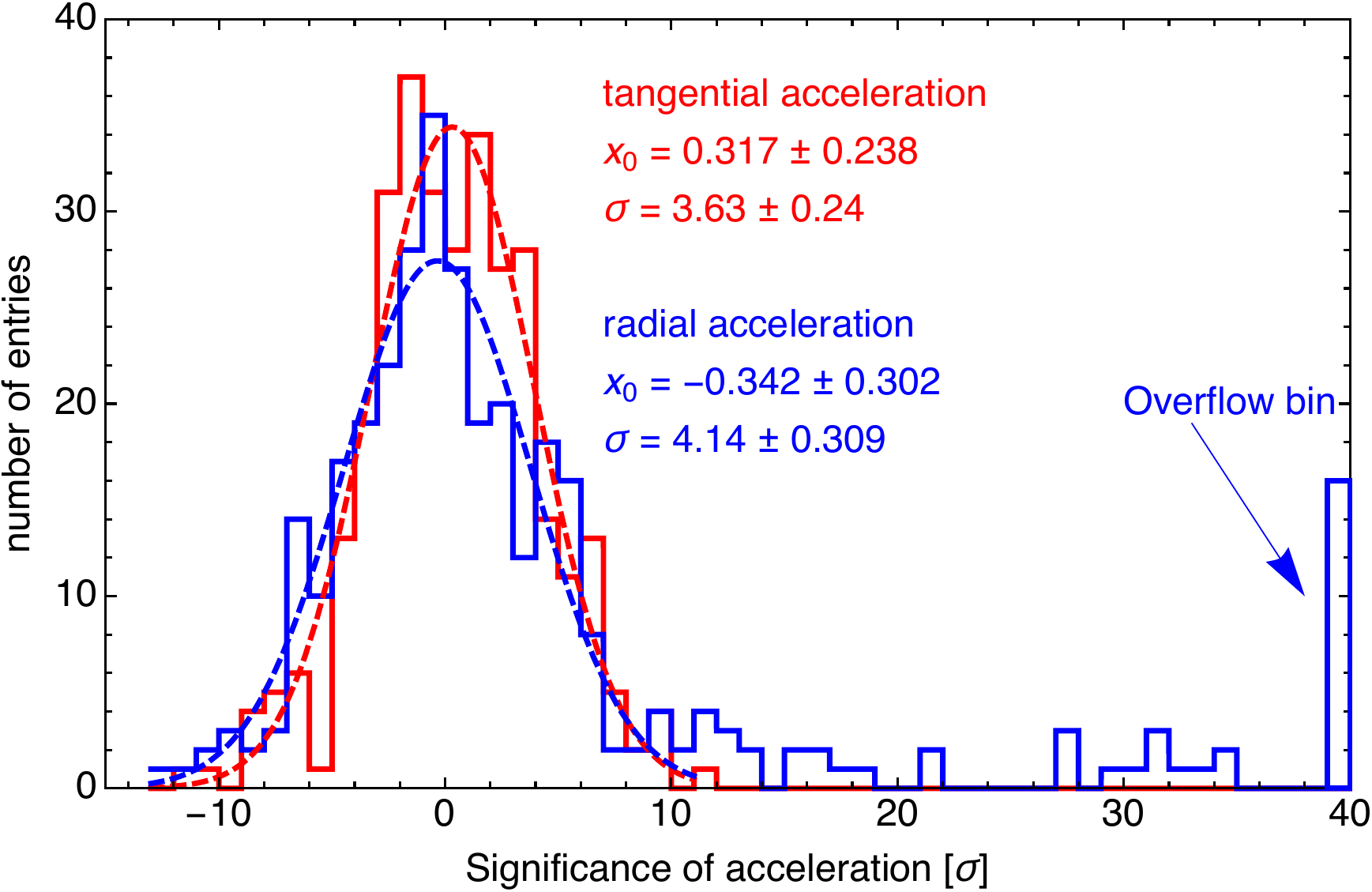}
\caption{Histograms of the significance of the accelerations. Red: in the tangential direction, i.e. perpendicular to the radial vector connecting the coordinate system origin and the star. Blue: in the radial direction. The sample consists both of the reference and the S-stars we monitor regularly, cleaned for duplicates (some S-stars also serve as reference stars), outliers due to confusion, stars with too large orbital phase coverage for a second order polynomial description, and applied a cut of $m_K < 18$. The significance is defined as the value of the quadratic coefficient in a second order polynomial fit divided by its error (and can thus be negative). 16 stars with radial accelerations $ \ge 40 \sigma$ are placed in the overflow bin at the right side.}
\label{fig_acchisto}
\end{figure}

In figure~\ref{fig_acchisto} we show histograms of the significances $\sigma$ of the measured accelerations. The significance is defined as the value of the coefficient of the second order term in a polynomial fit divided by its error, where we have rotated the coordinates of each star such that one axis points radially toward the position of Sgr~A*, and the other tangentially. The sample consists both of the reference stars and the S-stars we monitor regularly. We cleaned for duplicates (some S-stars also serve as reference stars), outliers due to confusion, stars with large orbital phase coverage like S2 (for which a second order polynomial is not a good description of the orbit), and we applied a cut of $m_K < 18$ (10~stars). The histogram for the radial accelerations shows an excess towards positive significances, as expected due to the gravitational force of the MBH, while the tangential accelerations can be described by a Gaussian distribution. The histograms also show that only at $\approx 8\sigma$ one can be reasonably sure that an acceleration is real. Given the small number of stars we are dealing with, we visually inspect each star individually and also check that an acceleration does not violate by more than $3\sigma$ the maximum value it can have given the 2D distance of the star from Sgr~A*.

Beyond S1 and S2 our data set contains 15 more stars that might be useful for constraining mass and distance. Table~\ref{tab_sstars_1} in the appendix identifies these stars by counting how many dynamical quantities are measured for each. They carry the label 'yes' in the column 'distance constraint'. The number of dynamical quantities is $N = 4 + (p-1) + (q+1)$, where $p$ is the polynomial order needed to describe the astrometry, $q$ the order needed for the radial velocity, and $q=-1$ for stars with no radial velocity. We demand $N \geq 8$ and $q \ge 0$. This yields 14 stars. In principle, for each of these stars we can fit an orbit keeping the potential free und using the coordinate system priors. In practice, the constraint might be (very) weak or biased. An example for a weak constraint is S38. Its inclination is almost face-on, such that the inclination-distance degeneracy discussed in section~\ref{sec_s1} is severe. An example for a star being too confused is S175. We do cover its percienter passage, but we unfortunately have no unbiased measurements. 

In addition, also stars without radial velocity but large phase coverage like S55 (S0-102 in \cite{2012Sci...338...84M}) can be useful for a multi-star fit, since they still constrain $\mu$ and the position of the central mass. S55 is the only such star in our sample.
We selected thus the following 17 stars for a multi-star fit: S2, S1, S4, S8, S9, S12, S13, S14, S17, S18, S19, S21, S24, S31, S38, S54, and S55.

\subsubsection{Multi-star fit}
\begin{figure*}
\centering
\includegraphics[width=0.66\linewidth]{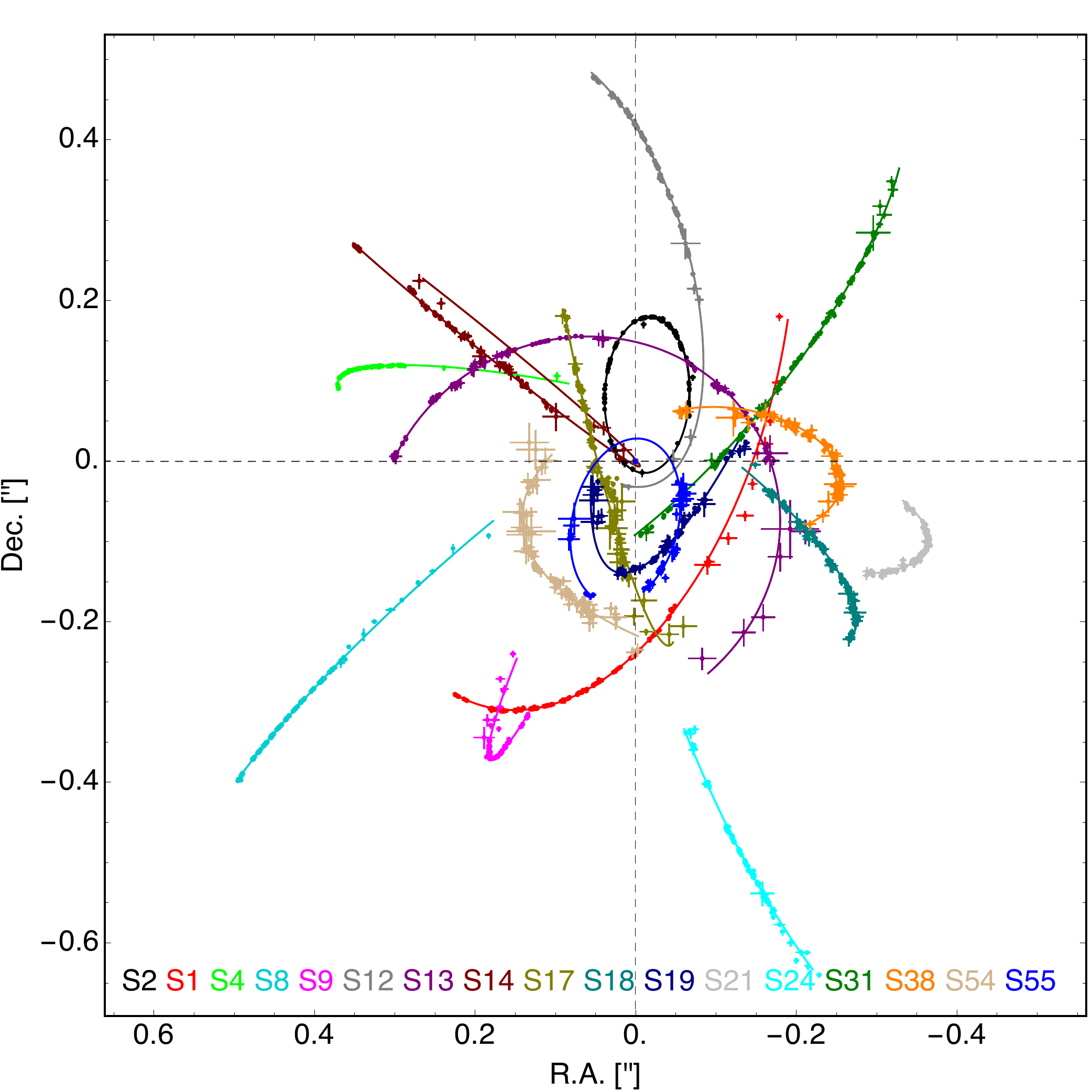}
\includegraphics[width=0.33\linewidth]{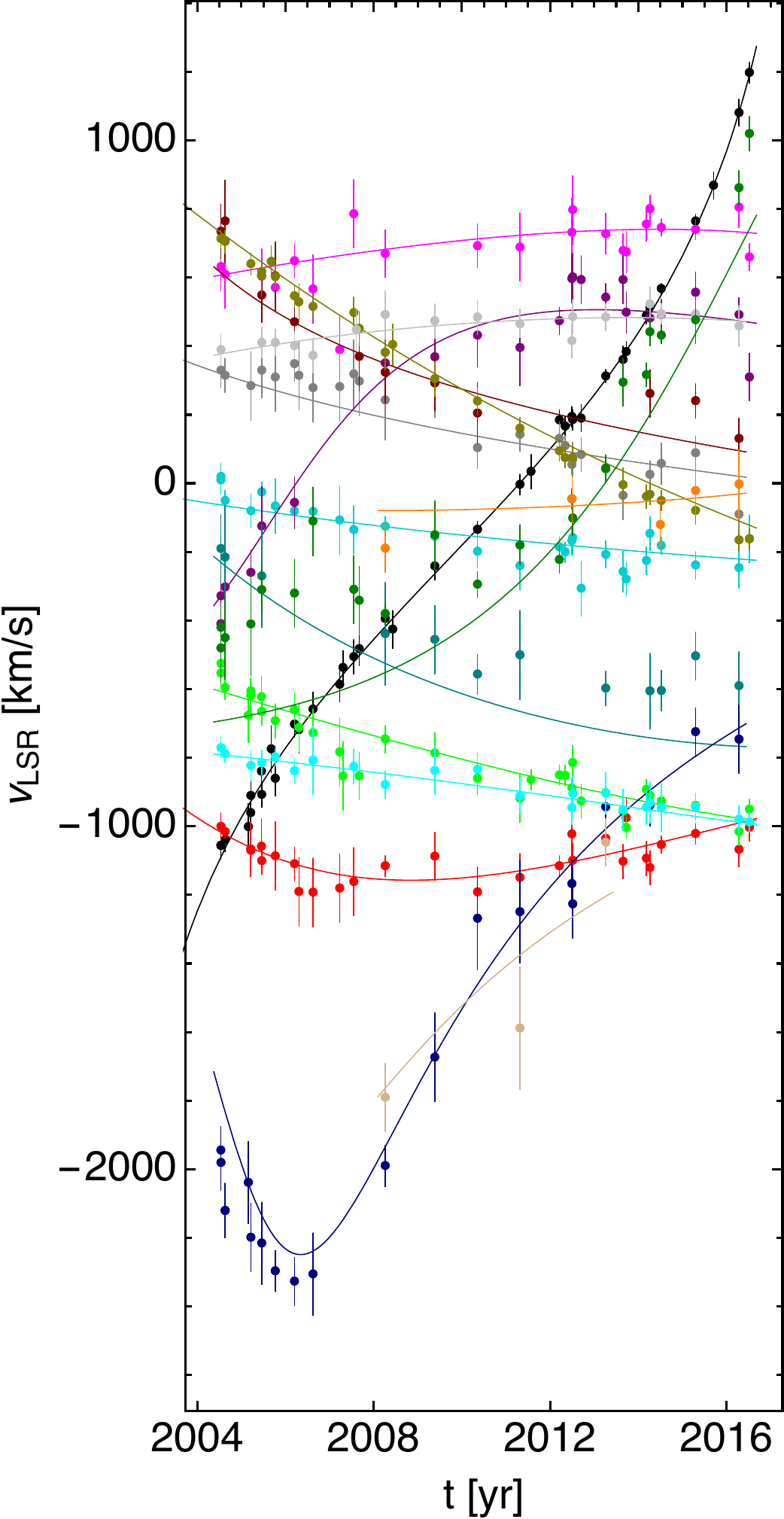}
\caption{Left: The astrometric data for the 17 stars used for the multi-star fit, shown together with the best-fitting orbits from the multi-star fit (solid lines). The dashed lines mark the position of the mass in this fit. Right: The radial velocity data of the sample, omitting the S2 data before 2004. The color coding is the same as for the left panel. Also, for S55 we don't have any radial velocity information. The solid lines give the best fitting orbits. }
\label{fig_plot17}
\end{figure*}

Fitting 17 stars simultaneously is a problem with 109 free parameters, for five of which (the coordinate system) we have prior information. Section~\ref{sec_artoffitting} (appendix) explains how we deal with this high-dimensional fit. Before starting the fit we have rescaled the error bars for each star individually such that it yields a reduced $\chi^2$ of 1 when fit alone in the fixed S2 potential. This gives each star a weight in the fit corresponding to the number of data points it contributes. The fit converges at the following distance estimate (row 9 in table~\ref{tab_fit_s2}):
\begin{figure}
\centering
\includegraphics[width=0.9\linewidth]{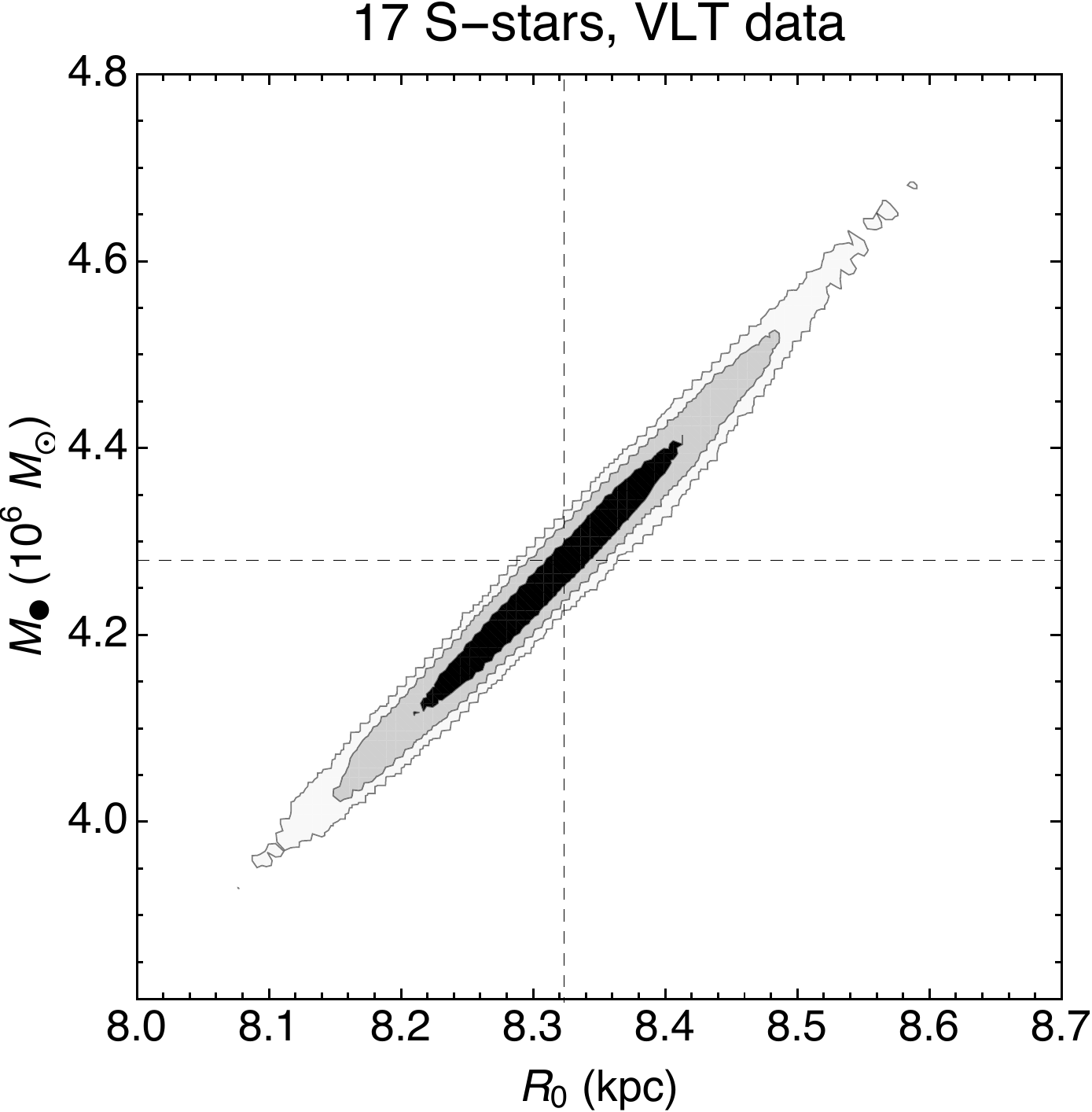}
\caption{Combined constraint on mass $M$ and distance $R_0$ using the multi-star fit. Note the different axes scale compared to figures~\ref{fig_massdist_s2} and~\ref{fig_bootstrap}.}
\label{fig_fit17}
\end{figure}

\begin{eqnarray}
M &=& 4.280 \pm 0.103 \times 10^6 \, M_\odot \nonumber \\ 
R_0 &=& 8.323 \pm 0.070  \, \mathrm{kpc} \,\,.
\label{eq_multistarfit}
\end{eqnarray}

The orbital data and the best fitting orbits for the multi-star fit are shown in fig.~\ref{fig_plot17} and the resulting parameter constraints for $R_0$ and $M$ in figure~\ref{fig_fit17}.

\subsubsection{Systematic errors}
\label{sec_syserr_multi}

The multi-star fit is dominated by the S2 data. Therefore, we base our estimate of the systematic error on the one for S2. We only replace the bootstrapping error from the S2 assessment by the weighted mean square deviation 
using those stars, for which we get a good constraint on $R_0$, i.e. S2, S1, S9 and S13. The weighted mean square deviation is
\begin{equation}
\sigma^2 = \frac{\Sigma_\mathrm{stars} (R_{0,\mathrm{star}} - R_{0,\mathrm{multi-star}})^2 / (\Delta R_{0,\mathrm{star}})^2}{\Sigma_\mathrm{stars} \, 1/ (\Delta R_{0,\mathrm{star}})^2} \,\,.
\label{eqMu}
\end{equation}
It yields a contribution of $0.10\,$kpc to the systematic error, which is also the dominant one. Together with the other systematic error components that we take identical to S2 the total systematic error for the multi-star fit is $0.14\,$kpc.

In row 10 in table~\ref{tab_fit_s2} we also report the results of a multi-star fit when excluding S2, i.e. using the 16 other stars only. The results do not differ significantly from the full multi-star fit, so the results are not driven by the S2 data only, and the
difference is compatible with our estimate of the systematic error. The best fit value for $R_0$ in the 16-star fit is at $8.19\,$kpc, while the mean of the posterior distribution from the Markov chain is at $8.24\,$kpc, indicative that the best fit value is slightly biased to lower values. This would further reduce the small difference between the 16-star fit and our fiducial 17-star fit.

\subsection{Positions of flares from Sgr~A*}

An additional cross-check is possible using the flaring emission of Sgr~A*. The stellar orbits locate the mass in the infrared coordinate system through orbit fitting. The location of the radio source Sgr~A* in the infrared coordinate system is based on the SiO masers \citep{2015MNRAS.453.3234P}. The flares from Sgr~A* locate the source directly in the infrared coordinate system.

Our data set contains 88 images in which we can identify the variable, radiative counterpart of Sgr~A* in the near-infrared. Fitting the positions with a linear motion yields a velocity of 
$(v_\mathrm{R.A.},\, v_\mathrm{Dec.}) = (-48,\,-41 ) \pm (174,\, 332)\,\mu$as/yr, i.e. consistent with being at rest. The error bar has been rescaled to yield a reduced $\chi^2$ of 1. This level of accuracy is similar to the constraints on how well we can fix Sgr~A* in the infrared coordinate system \citep{2015MNRAS.453.3234P}. Given the zero motion, we can determine the mean position of the flares. It is $(\Delta \,\mathrm{R.A.},\, \Delta\,\mathrm{Dec.}) = (-0.73,\,0.77 ) \pm (0.57,\, 1.08)\,$mas. The errors are dominated by the variance of the individual data points, as one might expect at the most confused position in our field of view. 

The flares' positions thus are consistent with the emission emanating from the position of the radio source Sgr~A*.

\subsubsection{The position of the central mass}
\label{sec_masspos}
\begin{figure}
\centering
\includegraphics[width=0.9\linewidth]{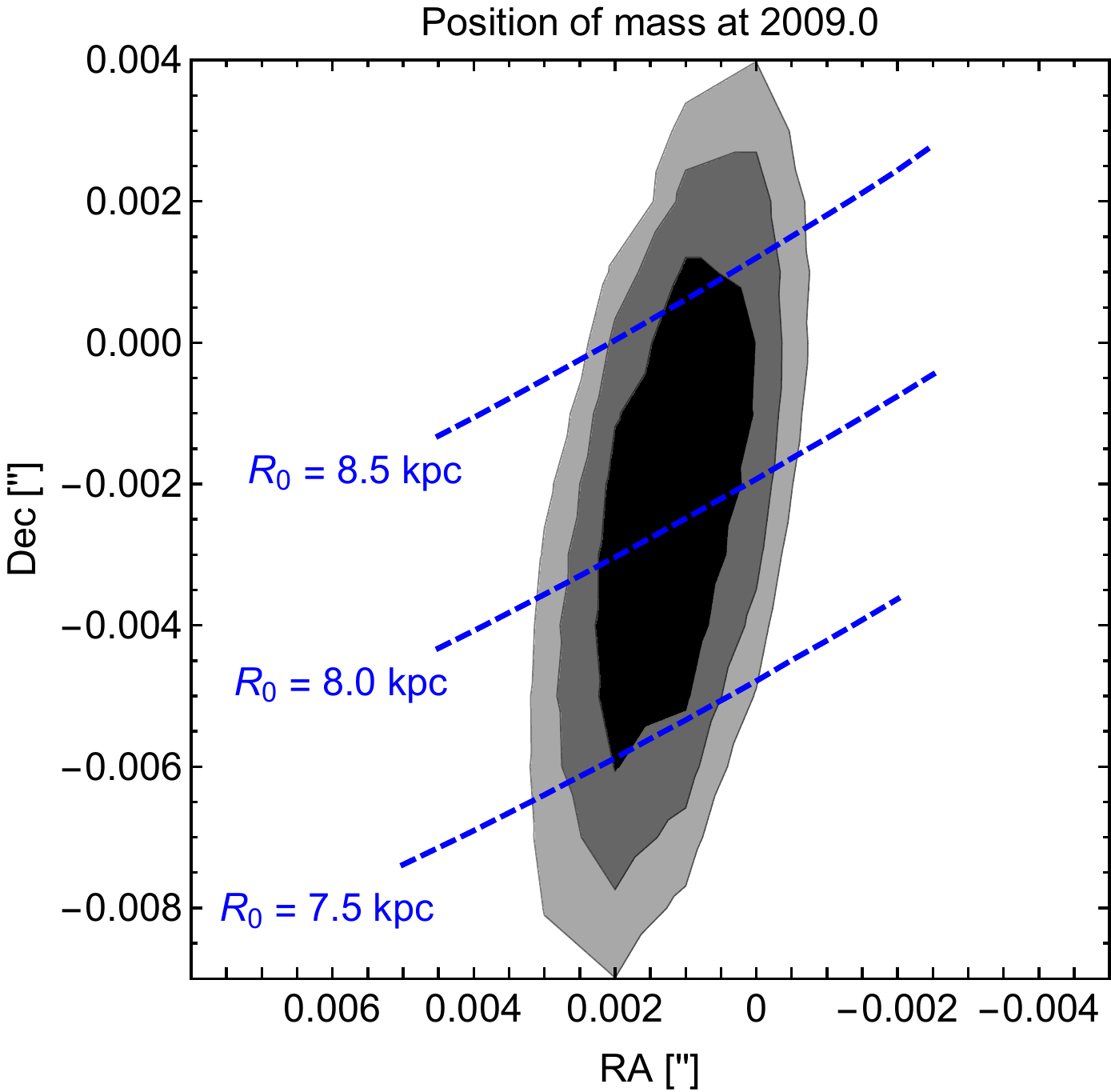}
\caption{The gray-toned areas give the levels of how significantly a fit using a fixed mass position differs from the best fit, shown as a function of the assumed, fixed position. The fits are rescaled such that the best fit has a reduced $\chi^2 = 1$, and the contours are drawn at levels of $\sqrt{2/\mathrm{d.o.f.}} = 0.0787$. One can read this plot as follows: Inside the black area the assumed position leads to a fit consistent at the 1-$\sigma$ level with the best fit with a free position. The epoch at which we fixed the mass is 2009.0, the epoch for which we have defined our reference frame. The blue, dashed lines mark the contours of the corresponding best-fitting values of $R_0$. }
\label{fig_masspos}
\end{figure}
The phase coverage of the S2 orbit with high accuracy measurements is incomplete so far: Mostly the northern half of the trajectory around apocenter has been mapped with the full accuracy. This means that the position of the central point mass is not yet determined as well as what we will achieve after the next pericenter passage. From the orbit's geometry one can expect that the current constraint in right ascension is better than in declination. 

Soon, a new type of measurement will assist the orbital monitoring: Astrometry from infrared interferometry with GRAVITY~\citep{2011Msngr.143...16E}. This novel instrument should be able to locate the position of the flaring emission to high precision with error bars below $100\,\mu$as. In order to prepare for such data, it is natural to ask in the context of this work: which externally measured positions of the mass are compatible with the current data set based on imaging? To that end we have fitted the combined S2 data with a set of orbits sampling different assumed positions of the central mass. We have kept the prior on the 2D velocity of the reference system unchanged for that. In figure~\ref{fig_masspos} we plot the reduced $\chi^2$ contours as a function of position indicating the significance levels defined by ${1\sigma = \sqrt{2/\mathrm{d.o.f.}}}$ \citep{2010arXiv1012.3754A}. The plot shows that our imaging data can accommodate mass positions in an area with an extent roughly $2\,\mathrm{mas} \times 6\,\mathrm{mas}$ in R.A. and declination. This is significantly larger than the statistical fit errors of the best fit. Note that the orientation of the area matches our expectation from the orbit geometry.

\subsection{Best estimate for $R_0$ and mass of Sgr~A*}

Our final, best estimate for $R_0$ and the mass of Sgr~A* is the Keplerian multi-star fit, for which we have:
\begin{eqnarray}
M &=& 4.28 \pm 0.10|_\mathrm{stat.} \pm 0.21|_\mathrm{sys} \times 10^6 \, M_\odot \nonumber \\ 
R_0 &=& 8.32 \pm 0.07|_\mathrm{stat.} \pm 0.14|_\mathrm{sys} \, \mathrm{kpc}   \,\,. 
\label{eq_best_est}
\end{eqnarray}
The scaling between mass and distance for that fit is
\begin{equation}
M(R_0) = (3.82 \pm 0.01) \times 10^6 M_\odot \times (R_0/8\,\mathrm{kpc})^{2.82} \,\, .
\end{equation}

\subsection{Comparison with Cluster data}

\begin{figure}
\centering
\includegraphics[width=0.9\linewidth]{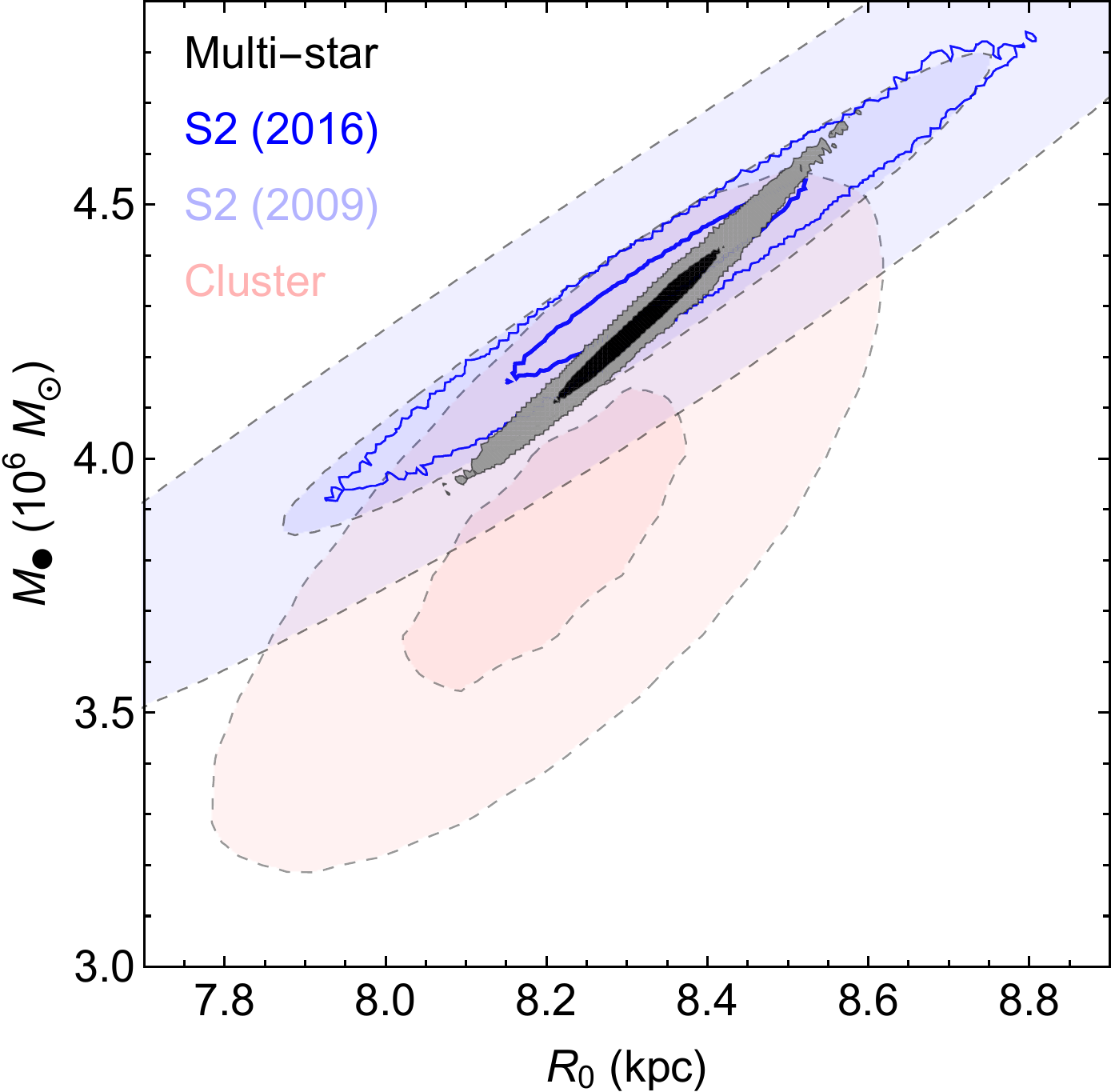}
\caption{Comparison of the results of measuring mass of and distance to Sgr~A* from stellar dynamical data, showing the respective $1\sigma$- and $3\sigma$-contours. The S2-based contours from \cite{2009ApJ...692.1075G} are given as light blue shaded areas, extending beyond the plot range. The S2 data as of 2016 (this work) yield the solid, darker blue contours. The 2016 $3\sigma$-contour matches roughly the 2009 $1\sigma$-contour. The dynamical modelling of the GC cluster data \citep{2015MNRAS.447..952C, 2016ApJ...821...44F} is shown as red shaded areas. Our multi-star fit yields the most stringent constraints, given by the black filled contours.}
\label{fig_combi_cluster}
\end{figure}

Our imaging and spectroscopy data set allows characterizing the dynamics of the larger-scale stellar cluster around Sgr~A*. From more than 10000 proper motions and more than 2500 radial velocities \citep{2016ApJ...821...44F}, \cite{2015MNRAS.447..952C} created a dynamical model of the cluster. Among the free parameters of the model are the central point mass $M$ and the distance $R_0$. This is a completely independent way of determining mass and distance, and we show the comparison of the constraints from S2, from the multi-star fit, and from the cluster data in figure~\ref{fig_combi_cluster}. The agreement is very satisfactory.

\section{Stellar orbits in the Galactic Center}

\begin{figure*}
\centering
\includegraphics[width=0.9\linewidth]{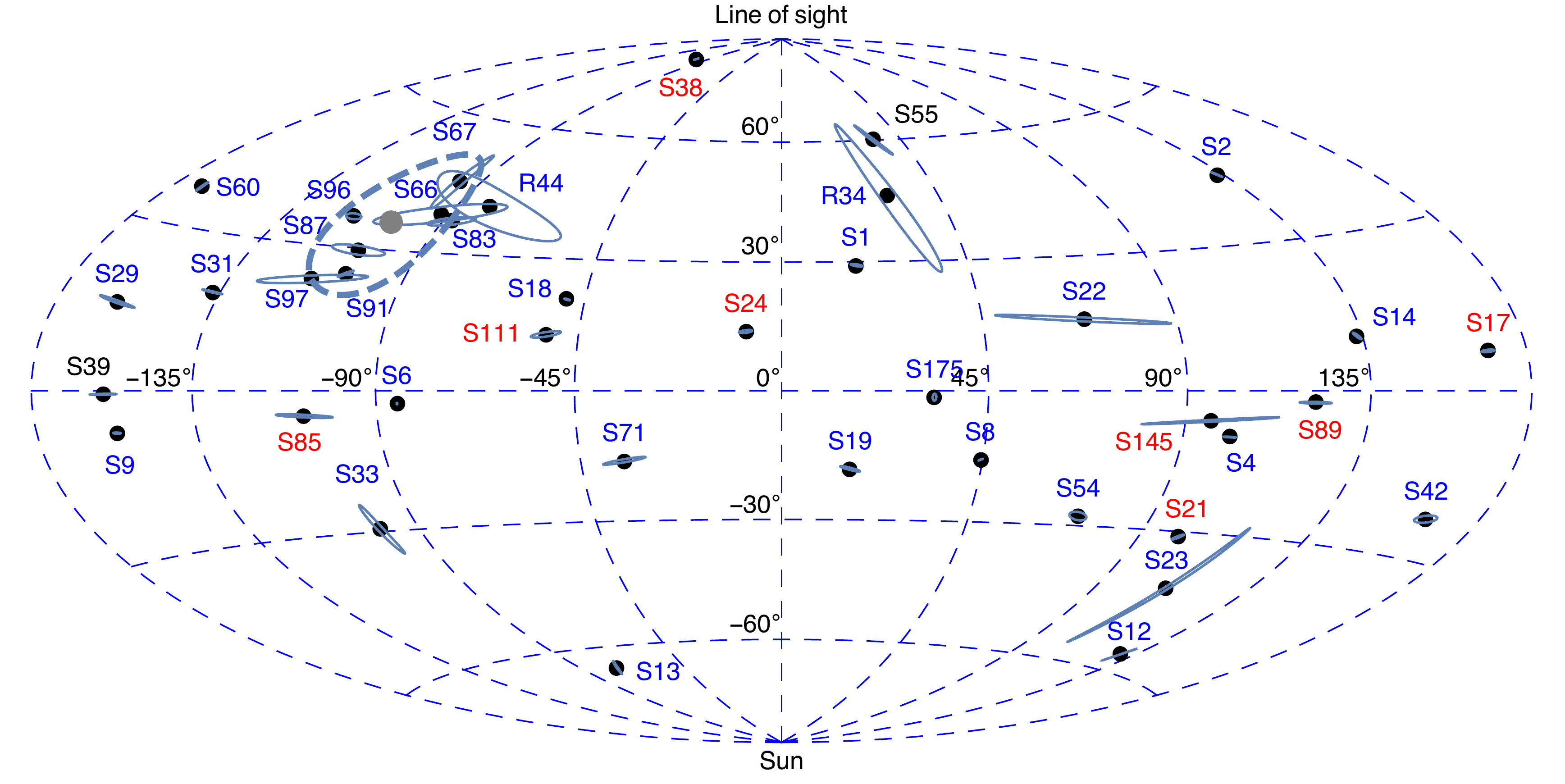}
\caption{Orientation of the orbital planes of those S-stars for which we were able to determine orbits. The vertical dimension corresponds to the inclination $i$ of the orbit and the horizontal dimension to the longitude of the ascending node $\Omega$. A star in a face-on, clockwise orbit relative to the line of sight, for instance, would be located at the top of the graph, while a star with an edge-on seen orbit would be located on the equator of the plot. The error ellipses correspond to the statistical 1-$\sigma$ fit errors only, thus the area covered by each is 39\% of the probability density function. Stars with an ambiguous inclination have been plotted at their more likely position. The stars S66, S67, S83, S87, S91, S96, S97 and R44 are members of the clockwise stellar disk 
\citep{2009ApJ...697.1741B, 2014ApJ...783..131Y} at ($ \Omega = 104^\circ$, $i = 126^\circ$) marked by the thick grey dot and the dashed line, indicating a disk thickness of $16^\circ$. The orbits of the other stars are oriented randomly. The color of the labels indicates the stellar type (blue for early-type stars, red for late-type stars).}
\label{fig_ang_mom}
\end{figure*}

\begin{figure*}
\centering
\includegraphics[width=0.45\linewidth]{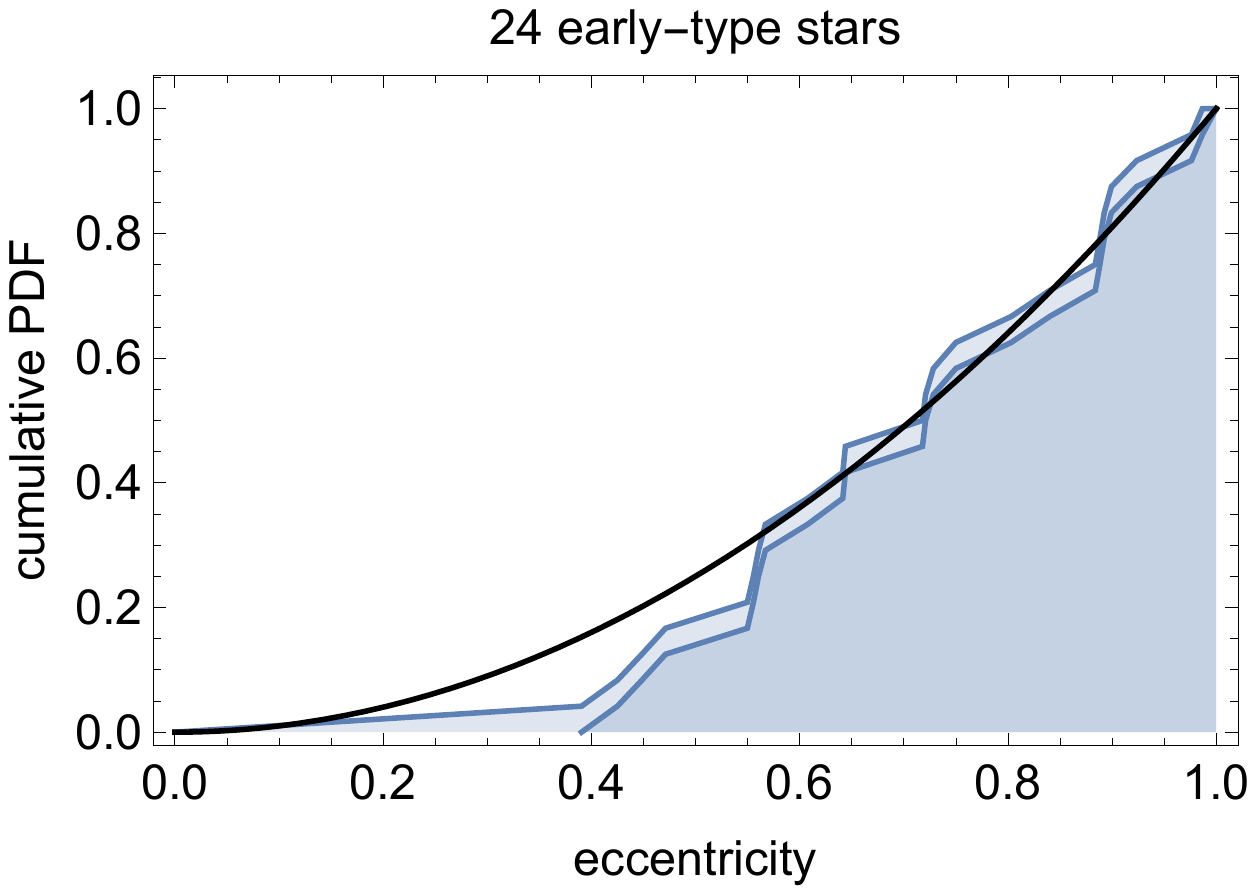}
\includegraphics[width=0.45\linewidth]{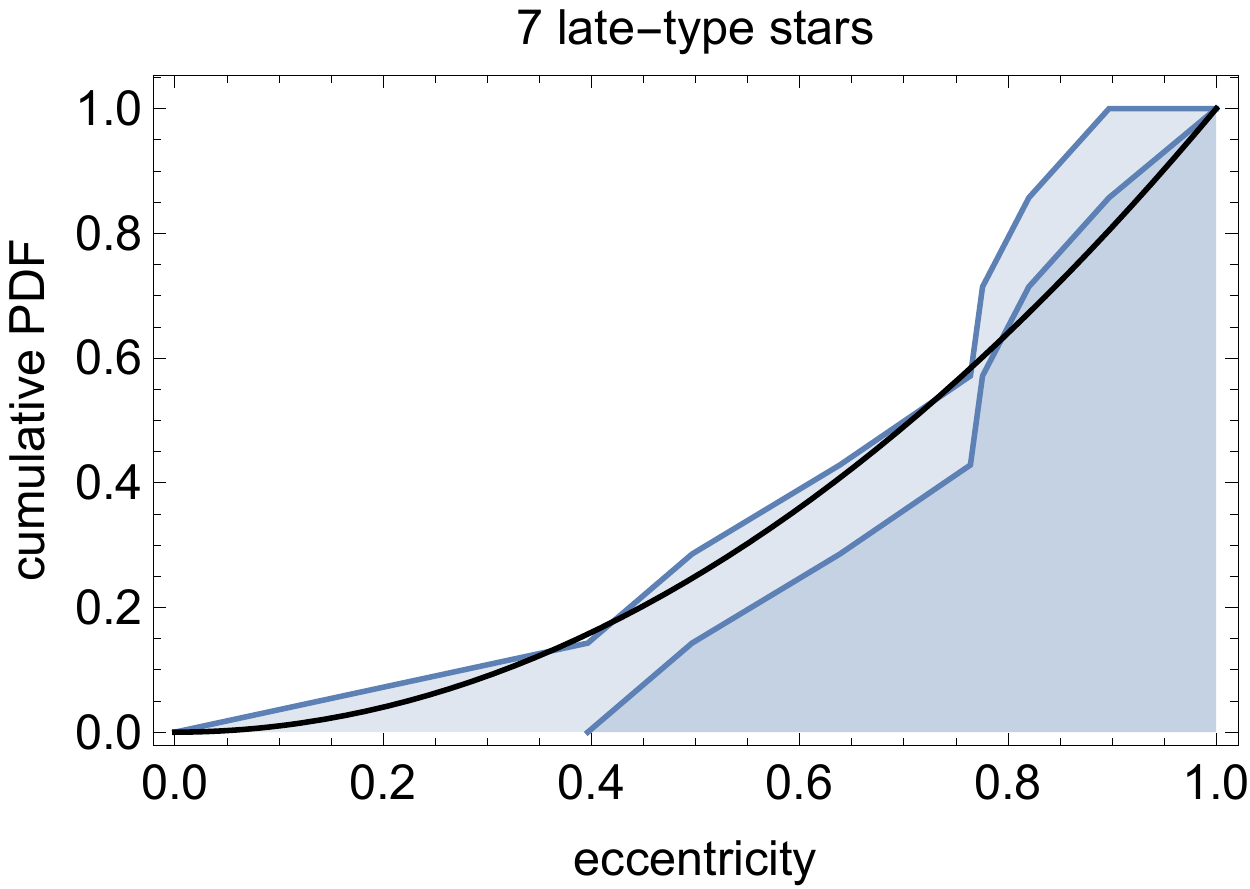}
\caption{Cumulative probability density function (PDF) for the eccentricities of the stars for which we have determined orbits. Left: The sample of 22 early-type stars, after exclusion of the eight stars which are identified as members of the stellar disk.  The two curves correspond to the two ways to plot a cumulative pdf, with values ranging either from 0 to (N - 1)/N or from 1/N to 1. The distribution is compatible with $n(e) \propto e$ (black line). Right: The same for the eight late-type stars for which we have determined orbits, excluding S111 which has a hyperbolic orbit, i.e. $e>1$.}
\label{fig_edist}
\end{figure*}

We have measured significant accelerations for 47 stars (table~\ref{tab_sstars_1}). We have demanded that the acceleration term be significant at least at the $8\sigma$ level, except for the stars S87, R34 and R44, which are bright and unconfused. For these, we trust the measured accelerations despite them being below the formal cut-off. For 40 of the stars we have sufficient information to determine an orbit, assuming the potential previously determined from S2 alone (see table~\ref{orbitsTable} in the appendix). Thirty of those are spectroscopically confirmed early-type stars, eight are late-type stars, and for S39 and S55 (the shortest period star, \cite{2012Sci...338...84M}) we don't know the spectral type. 

Compared to \cite{2009ApJ...692.1075G}, we don't give an orbital solution for S5 and S27 anymore. For S5, the significance of the acceleration is only marginal ($5\sigma$) and the star has been confused for the past years. Also, its radial velocity is constant at  $\approx 50\,$km/s, such that we cannot derive a reliable orbit. For the late-type star S27, it turned out that the acceleration previously quoted was due to a confusion event, and the radial velocity is constant at $\approx -120\,$km/s.

In figure~\ref{fig_ang_mom} we show the distribution of the angular momentum vectors for the 40 orbits. Eight of the orbits have an orientation that is compatible with the clockwise stellar disk \citep{2009ApJ...697.1741B, 2014ApJ...783..131Y} at ($ \Omega = 104^\circ$, $i = 126^\circ$): It is the same six stars as in \cite{2009ApJ...692.1075G}, S66, S67, S83, S87, S96, and S97, plus the stars S91 and R44. The rest of the stars have randomly oriented orbits. This solidifies the findings of \cite{2009ApJ...692.1075G}.

For the eccentricity distribution (figure~\ref{fig_edist}) we skip the eight disk stars, and the late-type star S111 that has a hyperbolic orbit. The remaining 22 early-type stars are thermally distributed ($n(e) \propto e$), and also for the seven late-type stars we don't see any hint of the distribution being different from thermal. This is somewhat different from the findings in \cite{2009ApJ...692.1075G}, where the eccentricity distribution appeared to be marginally more eccentric than thermal.  A Hills-mechanism origin \citep{1988Natur.331..687H} would make the S-stars relax towards the thermal distribution from the eccentric side, while in a disk-migration scenario the distribution would be approached from the less-than-thermal side. In our current data set, the eccentricity distribution is close to thermal and does not hold any such information on the formation scenario for the S-stars. But it does constrain the time scale on which the S-stars need to fully relax after they have been brought to the central arcsecond to be shorter than their life times.

The eight disk stars span only a limited range of lower eccentricities with $0.13 < e < 0.36$. The mean eccentricity is $<e> =0.26$, and the variance of the eight eccentricity values is 0.08. This is in perfect agreement with what \cite{2009ApJ...697.1741B} and \cite{2014ApJ...783..131Y} find.

\section{Summary}

We update the results from our long-term science program monitoring stellar orbits in the Galactic Center. Compared to our previous work \citep{2009ApJ...692.1075G} we can extend the time span covered with observations from 17 years to 25 years, and we implement the improved definition of the coordinate system from \cite{2015MNRAS.453.3234P}. Our main findings are:
\begin{itemize}
\item The statistical parameter uncertainties of the mass of and distance to Sgr~A* for an orbit fit of S2 have halved compared to \cite{2009ApJ...692.1075G}. Also the match between reconstructed position of the mass and expected position has improved.
\item If one applies the coordinate system priors, the star S1 yields a similarly good constraint on mass and distance as S2. 
\item From our sample of 47 stars for which we have measured accelerations, we can use 17 for a multi-star orbit fit. The latter yields our best estimates for mass and distance (eq.~\ref{eq_best_est}):
\begin{eqnarray}
R_0 &=& 8.32 \pm 0.07|_\mathrm{stat.} \pm 0.21|_\mathrm{sys} \, \mathrm{kpc} \nonumber \\ 
M &=& 4.28 \pm 0.10|_\mathrm{stat.} \pm 0.14|_\mathrm{sys} \times 10^6 \, M_\odot  \,\,.
\end{eqnarray}
The result agrees well with the statistical parallax estimated from the nuclear stellar cluster \citep{2015MNRAS.447..952C}. 
\item The positions of (i) the radio source Sgr~A*, (ii) the mass of $M\approx4 \times 10^6 M_\odot$, and (iii) the infrared-flares from Sgr~A* agree to $\approx 1\,$mas.
\item The distribution of eccentricities for the S-stars (after exclusion of eight stars that are members of the clockwise stellar disk) is completely compatible with a thermal distribution, for both early-type and late-type stars.
\item The orientation of orbital angular momenta of the same sample is random.
\item The eight disk stars have a mean eccentricity ${<e>} =0.26\pm 0.08$, and their mean distance to the disk direction from \cite{2009ApJ...697.1741B} is $12.0^\circ \pm 3.8^\circ$.
\end{itemize}
\section*{Acknowledgements}
SG and PP acknowledge the support from ERC starting grant No.~306311. RS was partially supported by iCore and ISF grants.
\FloatBarrier
~\newpage
\bibliographystyle{apj}
\bibliography{literature}
\FloatBarrier
~\newpage
\appendix
\section{Probability density function for the S2 orbit fit}
\begin{figure*}[h]
\centering
\includegraphics[width=0.95\linewidth]{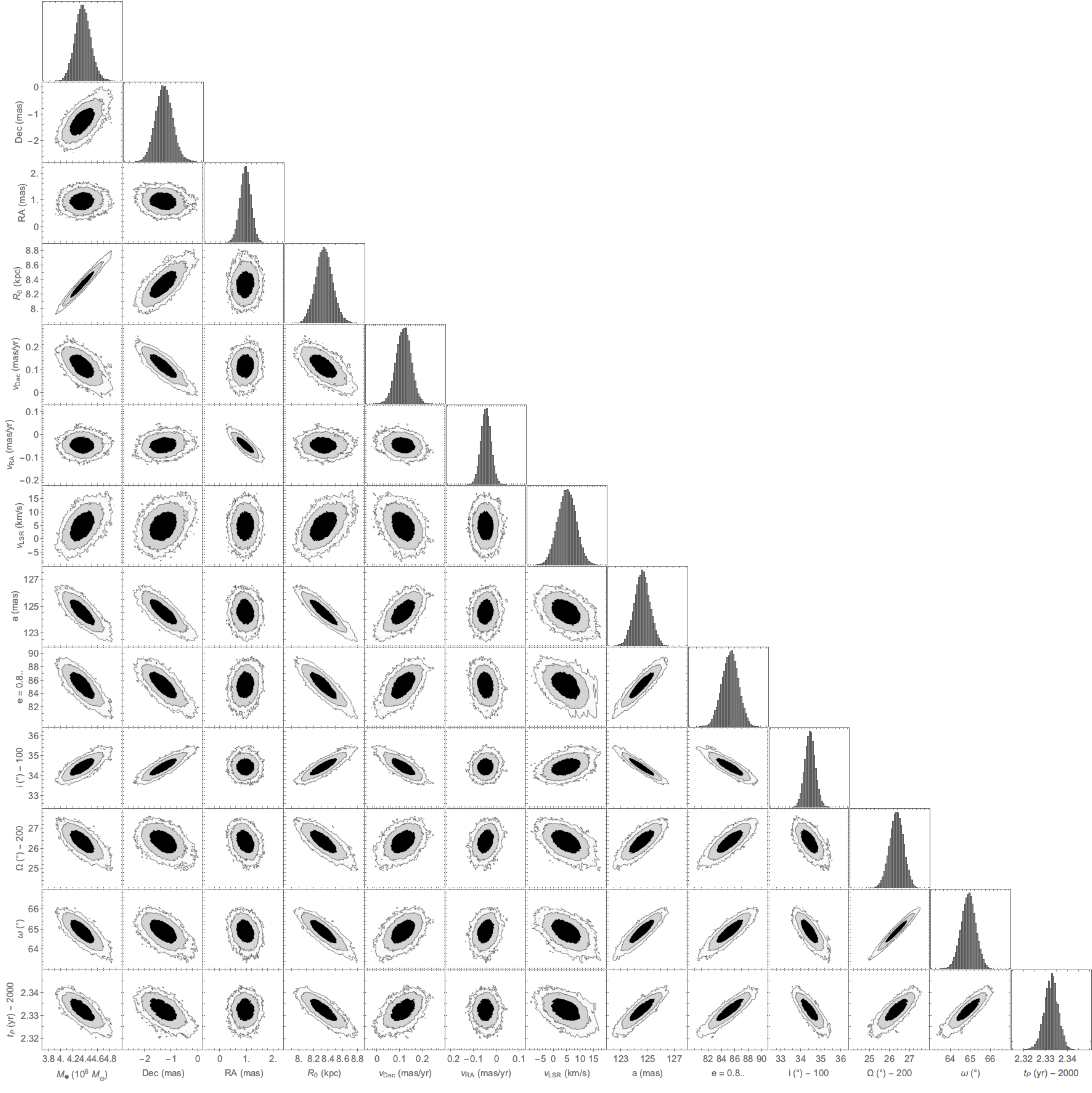}
\caption{Probability density function of 13 of the free parameters when fitting the combined S2 data set. To ease comparisons, parameters measured in identical units are plotted with identical axes lengths.}
\label{fig_full_mcmc}
\end{figure*}

In figure~\ref{fig_full_mcmc} we show the (almost) full probability density function for the combined S2 orbit fit. The four parameters describing the coordinate system mismatch between our and the Keck data are omitted for legibility. The probability density function (PDF) is compact and all parameters are well constrained, although there are significant correlations, such as between $M$ and $R_0$, between the coordinates of the central point mass and its velocity, or between inclination $i$ and semi major axis $a$.

~\pagebreak

\section{The art of fitting multiple orbits}
\label{sec_artoffitting}
Fitting 17 stars simultaneously is a problem with $7+17\times 6 = 109$ free parameters. We find that using a standard minimization routine takes prohibitively long for such a fit. Technically speaking, this is a result of the condition number of the correlation matrix being large, such that the numerical precision of standard double point arithmetics has a hard time calculating proper estimates of gradients in parameter space, or gradients being too shallow for the numerical precision. Hence, a different approach is needed. Our approach uses the following observations:
\begin{itemize}
\item The stars are treated as test particles. Hence, they don't interact with each other. The orbit parameters of two stars are not influencing each other directly, but they are all of course correlated with the potential parameters. This means that for a given potential, the $n\times 6$ square matrix separates into $n$ matrices of size $6\times 6$. For a given potential, one can optimize the orbital elements of the stars individually (and in parallel on multiple CPUs).
\item For a given set of orbital elements, six of the seven potential parameters are linear parameters (all except the mass). This means their optimum values can be calculated and don't need to be found iteratively. For a data set containing astrometry and radial velocity, the distance to the system ($R_0$) can be linearized, if the mass parameter $M$ is replaced by $\mu$.
\end{itemize}
Using that, the following algorithm yields a very good starting point for a minimization: 
\begin{itemize}
\item Use a one-dimensional search algorithm to minimize as a function of mass parameter $\mu$ the following $\chi^2_\mu$:
\item $\chi^2_\mu$ is the fixed point obtained by iterating $\chi^2$ until its value does not get smaller than a certain threshold anymore, when alternating the following two functions:
\begin{itemize}
\item Given the mass and the best estimates for the potential parameters at that mass optimize the parameters for all stars individually.
\item At the given mass and with the best orbital elements found, calculate the remaining six potential parameters that minimize $\chi^2$.
\end{itemize}
\end{itemize}
This procedure is not formally minimizing the 109 parameters simultaneously. However, empirically it yields a very good starting point, with which we can start a standard minimization routine. The latter finds a marginal improvement of the best $\chi^2$, however it stops at a position at which the error matrix still has a few, small negative Eigenvalues. Finding a further decrease in $\chi^2$ is impossible by the numerical precision employed. Instead, we use the best position in parameter space as a starting point for a Markov chain, which shows that the parameters found are very close to the global minimum, and which yields also the formal fit uncertainties. 

The above iterative procedure is efficient, if the individual minimizations per star are converging well. This is a practical concern, with which we can cope using the Thiele-Innes orbital elements \citep{2009ApJS..182..205W}. They allow formulating the problem in a convenient form for orbital fitting, namely a quadratic form of the $\chi^2$ in six parameters, which needs to be minimized with side constraints.

For a single orbit around a mass at a fixed position and an astrometry-only data set, four of the orbital elements can be linearized, and the iterative search only needs to deal with three parameters, namely period, pericenter time and eccentricity. Semi-major axis and the three angles can be calculated from the four linear Thiele-Innes elements, conventionally named ($A,\, B,\, G,\, F$). The search over period is of course equivalent to searching in mass space.

If multiple stars orbit the same mass, one cannot independently search the different periods $P_i$ and calculate the corresponding semi-major axes $a_i$, since for each pair $(P_i,\,a_i)$ the relation $G M = 4 \pi^2 a_i^3 / P_i^2$ needs to be fulfilled with the same mass $M$. This leads thus to side-constraints on the $a_i$ to be found. 

For a data set containing both astrometry and radial velocity data for a single star, one can only linearize two of the orbital elements ($a,\,\omega$). This uses two different Thiele-Innes elements, usually called ($C,\, H$). It means, one needs to search the four other orbital elements and the mass. If one has multiple stars with astrometric and radial velocity data, one gets again a side constraint on the semi-major axis $a$, such that one is left with only one linear parameter, $\omega$. The search space has not decreased much thus.

More convenient for the case of a multi-star fit with both astrometry and radial velocity data is a formulation in terms of ($A,\, B,\,C,\, G,\, F,\, H$), for which one gets a quadratic form of $\chi^2$, but which needs to be minimized with three three side constraints. This determines the three orbital angles in a robust way, and one needs to search the parameters space of $a$, $e$ and $t_p$ per star.
 
 ~\\
 ~\\
 ~\\
 ~\\
 ~\\
 ~\\
 ~\\
 ~\\
 ~\\
~\pagebreak

\section{Tabular overview of stars with measured accelerations}

\begin{table*}[h]
\caption{\label{tab_sstars_1} Table of stars for which we have measured a reliable acceleration}
{\scriptsize
\begin{center}
\begin{tabular}{l|ccccccccc|l}
star & polynom.& polynom. &N& signific. &phase & distance & period & $m_\mathrm{K}$ &spectral& comment\\
& order & order &  & accel. &coverage& constraint & [yr] & &type&\\
& astrometry & $v_\mathrm{LSR} $&& $[\sigma]$  & [rad]  & & && \\
\hline 
S1 & 4 & 2 &10& 127 &2.89 & yes & 166 & 14.8 & e & \\
S2 & $\infty$ & $\infty$ & $\infty$ & $\infty$ &6.86 & yes &16.0 & 14.1&e\\
S4 & 3 & 2 &9& 73 &1.08 & yes & 77 & 14.6& e\\
S6 & 2 &0& 6& 11&0.08& no &  192 &15.4& e\\
S8 & 3 &2 & 9&124 &0.60 & yes &   93 &14.4&e\\
S9 & 3 &1& 8&145 &0.88 & yes &  51  &15.2&e\\
S12 & $5$ & 1&10 &73&3.25 & yes &59 &15.6&e\\
S13 & 5 & 2 & 11 &56&4.47 & yes & 49 &16.1&e\\
S14 & $4$ & 1 &9&$\infty$&6.04 & yes &55 & 15.8&e\\
S17 & 3 & 2 & 9&0 &1.99 & yes & 77 &  16.0&l\\
S18 & 2 & 2 & 8&73&1.05 & yes &42 & 17.1& e\\
S19 & 3 & 2 &9&23&3.04  & yes  &135 &16.9&e\\
S21 & 3 & 1 &8&167&0.54 & yes &37 &16.9&l\\
S22 & 2 & 0 &6&11&0.46 & no &  540 &16.6&e\\
S23 & 2 & 0 &6&18&0.62 & no &  46 &17.9&e\\
S24 & 3 & 1 &8&34&0.52 & yes &  331 &15.8&l\\
S28 & 2 & N.A. & 5 &30&- & no & -& 17.3&-\\
S29 & 3 & 0 &7 &31&0.51 & no & 103  &16.9&e\\
S31 & 3 & 2 & 9 &108&1.92 & yes &  107 &16.0&e\\
S33 & 2 & 1 & 7&10&0.26 & no &192 &16.2&e\\
S38 & 3 & 1 &8&60&1.20&  yes & 19  & 17.2&l&$3^\mathrm{rd}$ order for astrometry marginal\\
S39 & 3 & N.A.&7 &35&0.54 & no &81 &17.4&-&$3^\mathrm{rd}$ order for astrometry marginal\\
S42 & 2 & 0 &6 &31&1.12 & no & 335  &17.3&e \\
S44 & 2 & N.A.& 5& 22&-  & no &- & 17.8&-& \\
S48 & 2 & N.A. &5 &27&-  & no &- & 16.8&-& \\
S54 & 3 & 1&8 &21&2.08& yes &  143 &17.6&e\\
S55 & $\infty$ & N.A.&$\infty$  &$\infty$&5.96 & no & 12.8  &17.6&-\\
S58 & 2 & N.A. &5 &13&-   & no&-& 17.9&- \\
S60 & 3 & 0 &7&16& 0.43& no & 87& 16.9& e\\
S64 &3  & N.A.& 6 &31&0.63  &no &25 & 17.6 & -& heavily confused with PSF of S2\\
S66 & 2 & 0 &6&14& 0.19 &no  &664 &14.8&e & CW disk star\\
S67 & 2 & 0 &6&28& 0.25 & no & 431&12.2&e & CW disk star\\
S71 & 2 & 0 &6&27&0.14 & no & 346&16.2&e\\
S82 & 2 & N.A. &5 &19&-  &no& -& 15.3&-& heavily confused with PSF of S95\\
S83 & 2 & 0 &6&36 &0.44 & no&656&13.7&e & CW disk star\\
S85 & 2 & 0 & 6&12&0.14  & no &3575 &15.5&l & \\
S87 & 2 & 0 &6&5& 0.11 & no&1637&13.6&e & CW disk star, acceleration marginal\\
S89 & 2 & 0 &6&12&0.11  &no  &406 &15.3&l & \\
S91 & 2 & 0 &6&22&0.27 & no &958 &12.6&e&CW disk star\\
S96 & 2 & 0 &6&16& 0.17 &no  & 662&10.5&e& CW disk star\\
S97 & 2 & 0 &6&17& 0.18 &no  & 1273&10.6&e& CW disk star\\
S111 &2 & 0&6&31&0.30 &no  & $\infty$&14.2&l& hyperbolic orbit\\
S145 & 2& 0 &6& 16 & 0.08  & no & 426 & 15.1 & l &\\
S146 & 2& N.A. & 5& 13 & -  & no & - & 15.9 & - &\\
S175 & $\infty$ & $\infty$ &$\infty$ &$\infty$&5.90 & no* & 96  &17.1&e& *too confused at all times\\
R34  & 2 & 0 &6&7&0.05 & no & 867&12.9&e& \\
R44  & 2 & 0 &6&7&0.10& no &2700 &13.5&e& CW disk star\\
\hline
\end{tabular}
\end{center}
}
\end{table*}

\FloatBarrier
\newpage
\section{Orbital elements for the stellar orbits}

\begin{table}[!ht]
\caption{Orbital parameters of the 40 stars for which we were able to determine orbits. The parameters were determined in the potential as obtained from the combined S2 data set,
the errors quoted in this table are the formal fit errors after rescaling them such that the reduced $\chi^2=1$ and including the uncertainties from
the potential. The last three columns give the spectral type ('e' for early-type stars, 'l' for late-type stars), the K-band magnitude, and the global rescaling factor
for that star. S111 formally has a negative semi major axis, indicative for a hyperbolic orbit with $e>1$.}
\label{orbitsTable} 
{\scriptsize
\begin{tabular}{lcccccccccc}
Star& $a$['']& $e$& $i\,[^\circ]$& $\Omega\,[^\circ]$& $\omega\,[^\circ]$& $t_P$[yr]& $T$[yr]&Sp&$m_K$&r\\
\hline
S1&$0.595\pm0.024$&$0.556\pm0.018$&$119.14\pm0.21$&$342.04\pm0.32$&$122.3\pm1.4$&$2001.80\pm0.15$&$166.0\pm5.8$&e&14.7&1.75\\
S2&$0.1255\pm0.0009$&$0.8839\pm0.0019$&$134.18\pm0.40$&$226.94\pm0.60$&$65.51\pm0.57$&$2002.33\pm0.01$&$16.00\pm0.02$&e&13.95&1.13\\
S4&$0.3570\pm0.0037$&$0.3905\pm0.0059$&$80.33\pm0.08$&$258.84\pm0.07$&$290.8\pm1.5$&$1957.4\pm1.2$&$77.0\pm1.0$&e&14.4&1.25\\
S6&$0.6574\pm0.0006$&$0.8400\pm0.0003$&$87.24\pm0.06$&$85.07\pm0.12$&$116.23\pm0.07$&$2108.61\pm0.03$&$192.0\pm0.17$&e&15.4&1.58\\
S8&$0.4047\pm0.0014$&$0.8031\pm0.0075$&$74.37\pm0.30$&$315.43\pm0.19$&$346.70\pm0.41$&$1983.64\pm0.24$&$92.9\pm0.41$&e&14.5&1.18\\
S9&$0.2724\pm0.0041$&$0.644\pm0.020$&$82.41\pm0.24$&$156.60\pm0.10$&$150.6\pm1.0$&$1976.71\pm0.92$&$51.3\pm0.70$&e&15.1&1.65\\
S12&$0.2987\pm0.0018$&$0.8883\pm0.0017$&$33.56\pm0.49$&$230.1\pm1.8$&$317.9\pm1.5$&$1995.59\pm0.04$&$58.9\pm0.22$&e&15.5&2.37\\
S13&$0.2641\pm0.0016$&$0.4250\pm0.0023$&$24.70\pm0.48$&$74.5\pm1.7$&$245.2\pm2.4$&$2004.86\pm0.04$&$49.00\pm0.14$&e&15.8&3.25\\
S14&$0.2863\pm0.0036$&$0.9761\pm0.0037$&$100.59\pm0.87$&$226.38\pm0.64$&$334.59\pm0.87$&$2000.12\pm0.06$&$55.3\pm0.48$&e&15.7&2.16\\
S17&$0.3559\pm0.0096$&$0.397\pm0.011$&$96.83\pm0.11$&$191.62\pm0.21$&$326.0\pm1.9$&$1991.19\pm0.41$&$76.6\pm1.0$&l&15.3&3.00\\
S18&$0.2379\pm0.0015$&$0.471\pm0.012$&$110.67\pm0.18$&$49.11\pm0.18$&$349.46\pm0.66$&$1993.86\pm0.16$&$41.9\pm0.18$&e&16.7&2.28\\
S19&$0.520\pm0.094$&$0.750\pm0.043$&$71.96\pm0.35$&$344.60\pm0.62$&$155.2\pm2.3$&$2005.39\pm0.16$&$135\pm14$&e&16.&2.57\\
S21&$0.2190\pm0.0017$&$0.764\pm0.014$&$58.8\pm1.0$&$259.64\pm0.62$&$166.4\pm1.1$&$2027.40\pm0.17$&$37.00\pm0.28$&l&16.9&1.60\\
S22&$1.31\pm0.28$&$0.449\pm0.088$&$105.76\pm0.95$&$291.7\pm1.4$&$95\pm20$&$1996.9\pm10.2$&$540\pm63$&e&16.6&2.78\\
S23&$0.253\pm0.012$&$0.56\pm0.14$&$48.0\pm7.1$&$249\pm13$&$39.0\pm6.7$&$2024.7\pm3.7$&$45.8\pm1.6$&e&17.8&2.08\\
S24&$0.944\pm0.048$&$0.8970\pm0.0049$&$103.67\pm0.42$&$7.93\pm0.37$&$290\pm15$&$2024.50\pm0.03$&$331\pm16$&l&15.6&1.54\\
S29&$0.428\pm0.019$&$0.728\pm0.052$&$105.8\pm1.7$&$161.96\pm0.80$&$346.5\pm5.9$&$2025.96\pm0.94$&$101.0\pm2.0$&e&16.7&3.32\\
S31&$0.449\pm0.010$&$0.5497\pm0.0025$&$109.03\pm0.27$&$137.16\pm0.30$&$308.0\pm3.0$&$2018.07\pm0.14$&$108.\pm1.2$&e&15.7&3.16\\
S33&$0.657\pm0.026$&$0.608\pm0.064$&$60.5\pm2.5$&$100.1\pm5.5$&$303.7\pm1.6$&$1928\pm12$&$192.0\pm5.2$&e&16.&2.21\\
S38&$0.1416\pm0.0002$&$0.8201\pm0.0007$&$171.1\pm2.1$&$101.06\pm0.24$&$17.99\pm0.25$&$2003.19\pm0.01$&$19.2\pm0.02$&l&17.&2.48\\
S39&$0.370\pm0.015$&$0.9236\pm0.0021$&$89.36\pm0.73$&$159.03\pm0.10$&$23.3\pm3.8$&$2000.06\pm0.06$&$81.1\pm1.5$& &16.8&3.27\\
S42&$0.95\pm0.18$&$0.567\pm0.083$&$67.16\pm0.66$&$196.14\pm0.75$&$35.8\pm3.2$&$2008.24\pm0.75$&$335\pm58$&e&17.5&1.65\\
S54&$1.20\pm0.87$&$0.893\pm0.078$&$62.2\pm1.4$&$288.35\pm0.70$&$140.8\pm2.3$&$2004.46\pm0.07$&$477\pm199$&e&17.5&2.60\\
S55&$0.1078\pm0.0010$&$0.7209\pm0.0077$&$150.1\pm2.2$&$325.5\pm4.0$&$331.5\pm3.9$&$2009.34\pm0.04$&$12.80\pm0.11$& &17.5&1.61\\
S60&$0.3877\pm0.0070$&$0.7179\pm0.0051$&$126.87\pm0.30$&$170.54\pm0.85$&$29.37\pm0.29$&$2023.89\pm0.09$&$87.1\pm1.4$&e&16.3&1.65\\
S66&$1.502\pm0.095$&$0.128\pm0.043$&$128.5\pm1.6$&$92.3\pm3.2$&$134\pm17$&$1771\pm38$&$664\pm37$&e&14.8&1.70\\
S67&$1.126\pm0.026$&$0.293\pm0.057$&$136.0\pm1.1$&$96.5\pm6.4$&$213.5\pm1.6$&$1705\pm22$&$431\pm10$&e&12.1&1.43\\
S71&$0.973\pm0.040$&$0.899\pm0.013$&$74.0\pm1.3$&$35.16\pm0.86$&$337.8\pm4.9$&$1695\pm21$&$346\pm11$&e&16.1&1.87\\
S83&$1.49\pm0.19$&$0.365\pm0.075$&$127.2\pm1.4$&$87.7\pm1.2$&$203.6\pm6.0$&$2046.8\pm6.3$&$656\pm69$&e&13.6&1.82\\
S85&$4.6\pm3.30$&$0.78\pm0.15$&$84.78\pm0.29$&$107.36\pm0.43$&$156.3\pm6.8$&$1930.2\pm9.8$&$3580\pm2550$&l&15.6&1.50\\
S87&$2.74\pm0.16$&$0.224\pm0.027$&$119.54\pm0.87$&$106.32\pm0.99$&$336.1\pm7.7$&$611\pm154$&$1640\pm105$&e&13.6&1.38\\
S89&$1.081\pm0.055$&$0.639\pm0.038$&$87.61\pm0.16$&$238.99\pm0.18$&$126.4\pm4.0$&$1783\pm26$&$406\pm27$&l&15.3&1.16\\
S91&$1.917\pm0.089$&$0.303\pm0.034$&$114.49\pm0.32$&$105.35\pm0.74$&$356.4\pm1.6$&$1108\pm69$&$958\pm50$&e&12.2&1.33\\
S96&$1.499\pm0.057$&$0.174\pm0.022$&$126.36\pm0.96$&$115.66\pm0.59$&$233.6\pm2.4$&$1646\pm16$&$662\pm29$&e&10.&1.31\\
S97&$2.32\pm0.46$&$0.35\pm0.11$&$113.0\pm1.3$&$113.2\pm1.4$&$28\pm14$&$2132\pm29$&$1270\pm309$&e&10.3&1.22\\
S111&$-12.3\pm8.4$&$1.092\pm0.064$&$102.68\pm0.40$&$52.34\pm0.75$&$132.4\pm3.3$&$1947.7\pm4.5$&$N.A..$&l&13.8&0.97\\
S145&$1.12\pm0.18$&$0.50\pm0.25$&$83.7\pm1.6$&$263.92\pm0.94$&$185\pm16$&$1808\pm58$&$426\pm71$&l&17.5&1.46\\
S175&$0.414\pm0.039$&$0.9867\pm0.0018$&$88.53\pm0.60$&$326.83\pm0.78$&$68.52\pm0.40$&$2009.51\pm0.01$&$96.2\pm5.0$&e&17.5&2.72\\
R34&$1.81\pm0.15$&$0.641\pm0.098$&$136.0\pm8.3$&$330\pm19$&$57.0\pm8.0$&$1522\pm52$&$877\pm83$&e&14.&1.35\\
R44&$3.9\pm1.4$&$0.27\pm0.27$&$131.0\pm5.2$&$80.5\pm7.1$&$217\pm24$&$1963\pm85$&$2730\pm1350$&e&14.&1.11\\
\end{tabular}
}
\end{table}

\newpage
\FloatBarrier

\section{Laws of motion for stars without orbits}

\begin{table*}[!ht]
\caption{\label{motionsTable} Laws of motions of stars in the central arcsecond. We select stars, which have $r < 1.2"$, $m_K \le 17.5$, and which are not yet covered in table~\ref{orbitsTable}. We include also all stars for which we have measured a significant acceleration (table~\ref{tab_sstars_1}) but cannot give an orbital solution. For these stars, the law of motion is quadratic, for the others it is linear.}
{\scriptsize
\begin{tabular}{lcc|l|l}
Star& $m_K$ &$t_0$ & $\Delta\,\mathrm{R.A.} (\Delta t)$ in mas &$\Delta\,\mathrm{Dec.} (\Delta t)$ in mas\\
\hline
\hline
S5& 15.2 & 2009.02 &$(327.7 \pm 0.3) + (-5.66 \pm 0.08) \Delta t$&
 $(225.2 \pm 0.3) + (7.57 \pm 0.06) \Delta t$\\
\hline
S7& 15.3 & 2004.38 &$(516.4 \pm 0.1) + (-3.70 \pm 0.02) \Delta t$&
 $(-46.9 \pm 0.1) + (-2.90 \pm 0.02) \Delta t$\\
\hline
S10& 14.1 & 2005.42 &$(43.0 \pm 0.0) + (-5.04 \pm 0.01) \Delta t$&
 $(-374.7 \pm 0.1) + (3.04 \pm 0.01) \Delta t$\\
\hline
S11& 14.3 & 2004.38 &$(176.1 \pm 0.1) + (8.79 \pm 0.02) \Delta t$&
 $(-574.9 \pm 0.1) + (-5.58 \pm 0.02) \Delta t$\\
\hline
S20& 15.7 & 2009.94 &$(200.3 \pm 0.1) + (-5.04 \pm 0.05) \Delta t$&
 $(81.4 \pm 0.1) + (-5.63 \pm 0.04) \Delta t$\\
\hline
S25& 15.2 &2005.42 & $(-107.6 \pm 0.1) + (-2.87 \pm 0.02) \Delta t$&
 $(-426.6 \pm 0.1) + (1.27 \pm 0.02) \Delta t$\\
\hline
S26& 15.1 & 2005.42 &$(539.9 \pm 0.2) + (6.30 \pm 0.05) \Delta t$&
 $(439.4 \pm 0.1) + (0.85 \pm 0.02) \Delta t$\\
\hline
S27& 15.6 & 2005.42 &$(148.9 \pm 0.1) + (0.63 \pm 0.02) \Delta t$&
 $(531.5 \pm 0.1) + (3.07 \pm 0.02) \Delta t$\\
\hline
S28& 17.1 &2003.02 &  $(-13.5 \pm 0.4) + (4.81 \pm 0.17) \Delta t + (0.173 \pm 0.025) \Delta t^2$&
$(441.1 \pm 0.4) + (10.00 \pm 0.17) \Delta t + (-0.631 \pm 0.023) \Delta t^2$\\
\hline
S30& 14.3 &2004.38 & $(-556.9 \pm 0.0) + (1.20 \pm 0.01) \Delta t$&
 $(393.9 \pm 0.0) + (3.39 \pm 0.01) \Delta t$\\
\hline
S32& 16.6 &2009.39 & $(-339.5 \pm 0.1) + (-3.82 \pm 0.03) \Delta t$&
 $(-358.2 \pm 0.1) + (0.47 \pm 0.03) \Delta t$\\
\hline
S34& 15.5 & 2006.03 &$(339.3 \pm 0.1) + (9.34 \pm 0.03) \Delta t$&
 $(-459.3 \pm 0.1) + (3.76 \pm 0.03) \Delta t$\\
\hline
S35& 13.3 & 2004.38 &$(546.2 \pm 0.1) + (1.82 \pm 0.01) \Delta t$&
 $(-430.1 \pm 0.1) + (3.06 \pm 0.01) \Delta t$\\
\hline
S36& 16.4 &2008.53 & $(274.0 \pm 0.3) + (-0.61 \pm 0.07) \Delta t$&
 $(234.7 \pm 0.4) + (-1.67 \pm 0.10) \Delta t$\\
\hline
S37& 16.1 &2009.43 & $(309.9 \pm 0.3) + (-5.58 \pm 0.08) \Delta t$&
 $(424.7 \pm 0.2) + (9.89 \pm 0.06) \Delta t$\\
\hline
S41& 17.5 &2009.33 & $(-214.6 \pm 0.2) + (1.20 \pm 0.05) \Delta t$&
 $(-313.7 \pm 0.2) + (-2.22 \pm 0.06) \Delta t$\\
\hline
S44& 17.5 & 2010.49 &$(-127.0 \pm 0.6) + (-6.79 \pm 0.26) \Delta t + (0.333 \pm 0.056) \Delta t^2$&
 $(-273.7 \pm 0.5) + (-1.79 \pm 0.18) \Delta t + (0.957 \pm 0.041) \Delta t^2$\\
\hline
S45& 15.7 &2009.43 & $(169.3 \pm 0.2) + (-5.58 \pm 0.05) \Delta t$&
 $(-538.4 \pm 0.1) + (-4.55 \pm 0.03) \Delta t$\\
\hline
S46& 15.7 &2005.42 & $(246.5 \pm 0.2) + (0.44 \pm 0.03) \Delta t$&
 $(-556.6 \pm 0.3) + (4.66 \pm 0.05) \Delta t$\\
\hline
S47& 16.3 & 2008.03 &$(378.1 \pm 0.4) + (-2.84 \pm 0.18) \Delta t$&
 $(247.1 \pm 0.4) + (4.26 \pm 0.15) \Delta t$\\
\hline
S48& 16.6 &2009.43 & $(432.5 \pm 0.2) + (-2.27 \pm 0.06) \Delta t + (-0.242 \pm 0.014) \Delta t^2$&
 $(513.5 \pm 0.2) + (11.00 \pm 0.05) \Delta t + (-0.187 \pm 0.011) \Delta t^2$\\
\hline
S50& 17.2 &2009.43 &  $(-511.9 \pm 0.2) + (-1.83 \pm 0.06) \Delta t$&
$(-492.2 \pm 0.3) + (9.96 \pm 0.08) \Delta t$\\
\hline
S51& 17.4 & 2009.43 &$(-444.3 \pm 0.2) + (7.55 \pm 0.06) \Delta t$&
 $(-271.7 \pm 0.2) + (8.02 \pm 0.05) \Delta t$\\
\hline
S52& 17.1 & 2006.97 &$(198.7 \pm 0.6) + (7.80 \pm 0.27) \Delta t$&
 $(290.1 \pm 1.1) + (-8.43 \pm 0.46) \Delta t$\\
\hline
S53& 17.2 & 2007.64 &$(325.1 \pm 0.4) + (8.83 \pm 0.16) \Delta t$&
 $(510.6 \pm 0.3) + (8.50 \pm 0.14) \Delta t$\\
\hline
S56& 17.0 &2011.85 & $(35.8 \pm 0.8) + (-25.62 \pm 0.34) \Delta t$&
 $(146.4 \pm 0.5) + (-0.75 \pm 0.22) \Delta t$\\
\hline
S58& 17.4 &2009.43 &  $(-314.3 \pm 0.2) + (7.34 \pm 0.08) \Delta t + (0.172 \pm 0.017) \Delta t^2$&
$(-553.3 \pm 0.2) + (5.23 \pm 0.08) \Delta t + (0.141 \pm 0.017) \Delta t^2$\\
\hline
S65& 13.7 &2004.38 &  $(-769.3 \pm 0.0) + (2.39 \pm 0.01) \Delta t$&
$(-279.1 \pm 0.0) + (-1.48 \pm 0.01) \Delta t$\\
\hline
S68& 12.9 & 2004.38 &$(295.4 \pm 0.1) + (5.13 \pm 0.03) \Delta t$&
 $(771.7 \pm 0.2) + (2.80 \pm 0.03) \Delta t$\\
\hline
S69& 16.8 & 2007.42 &$(-16.6 \pm 0.2) + (-0.13 \pm 0.08) \Delta t$&
 $(762.9 \pm 0.4) + (1.73 \pm 0.13) \Delta t$\\
\hline
S70& 16.9 &2009.43 & $(-360.2 \pm 0.1) + (-3.16 \pm 0.03) \Delta t$&
 $(694.7 \pm 0.1) + (-4.08 \pm 0.03) \Delta t$\\
\hline
S72& 14.3 &2004.38 & $(-634.3 \pm 0.1) + (9.04 \pm 0.01) \Delta t$&
 $(-886.2 \pm 0.1) + (-5.74 \pm 0.01) \Delta t$\\
\hline
S73& 16.1 & 2008.00 &$(-350.9 \pm 0.1) + (-9.85 \pm 0.02) \Delta t$&
 $(-1034.3 \pm 0.1) + (-8.61 \pm 0.02) \Delta t$\\
\hline
S74& 16.9 &2009.43 &  $(-101.3 \pm 0.1) + (-0.49 \pm 0.03) \Delta t$&
$(-846.5 \pm 0.1) + (4.93 \pm 0.04) \Delta t$\\
\hline
S75& 17.1 & 2009.43 & $(-129.5 \pm 0.2) + (6.57 \pm 0.05) \Delta t$&
$(-726.7 \pm 0.1) + (1.20 \pm 0.04) \Delta t$\\
\hline
S76& 12.8 &2004.38 & $(339.1 \pm 0.1) + (-3.80 \pm 0.01) \Delta t$&
 $(-913.5 \pm 0.1) + (4.22 \pm 0.01) \Delta t$\\
\hline
S77& 15.8 &2009.46 & $(397.0 \pm 0.3) + (8.74 \pm 0.09) \Delta t$&
 $(-860.4 \pm 0.2) + (-7.66 \pm 0.07) \Delta t$\\
\hline
S78& 16.5 &2009.43 & $(382.9 \pm 0.2) + (-17.48 \pm 0.05) \Delta t$&
 $(-699.9 \pm 0.2) + (-7.61 \pm 0.07) \Delta t$\\
\hline
S79& 16.0 &2009.43 & $(646.6 \pm 0.2) + (0.19 \pm 0.06) \Delta t$&
 $(-530.5 \pm 0.2) + (2.13 \pm 0.06) \Delta t$\\
\hline
S80& 16.9 & 2009.43 &$(971.3 \pm 0.1) + (-4.91 \pm 0.05) \Delta t$&
 $(-337.6 \pm 0.2) + (4.77 \pm 0.06) \Delta t$\\
\hline
S81& 17.2 & 2010.52 &$(754.5 \pm 8.5) + (-7.12 \pm 2.51) \Delta t$&
 $(-482.2 \pm 4.1) + (0.25 \pm 1.19) \Delta t$\\
\hline
S82& 15.4 & 2009.43 & $(24.6 \pm 0.1) + (-7.69 \pm 0.03) \Delta t + (0.015 \pm 0.007) \Delta t^2$&
$(881.7 \pm 0.2) + (-16.04 \pm 0.05) \Delta t + (-0.211 \pm 0.011) \Delta t^2$\\
\hline
S84& 14.4 & 2005.42 &$(-1118.5 \pm 0.1) + (4.47 \pm 0.01) \Delta t$&
 $(-24.2 \pm 0.1) + (1.74 \pm 0.01) \Delta t$\\
\hline
S86& 15.5 & 2004.38 &$(-1034.7 \pm 0.6) + (-0.17 \pm 0.14) \Delta t$&
 $(207.8 \pm 0.5) + (-4.13 \pm 0.11) \Delta t$\\
\hline
S88& 15.8 &2004.38 & $(-1026.5 \pm 0.1) + (-3.94 \pm 0.02) \Delta t$&
 $(-535.9 \pm 0.1) + (-7.87 \pm 0.02) \Delta t$\\
\hline
S90&16.1 & 2009.43 & $(534.0 \pm 0.2) + (0.90 \pm 0.05) \Delta t$&
$(-972.8 \pm 0.1) + (0.56 \pm 0.03) \Delta t$\\
\hline
S92& 13.0 & 2009.39 &$(1039.0 \pm 0.0) + (5.86 \pm 0.01) \Delta t$&
 $(30.1 \pm 0.1) + (1.05 \pm 0.01) \Delta t$\\
\hline
S93& 15.6 & 2009.43 &$(1072.1 \pm 0.1) + (-2.83 \pm 0.03) \Delta t$&
 $(159.1 \pm 0.1) + (-2.50 \pm 0.05) \Delta t$\\
\hline
S94& 16.7 &2009.04 & $(-189.1 \pm 0.3) + (-9.78 \pm 0.10) \Delta t$&
 $(913.5 \pm 0.3) + (1.99 \pm 0.12) \Delta t$\\
\hline
S98&15.6 &2005.42 & $(-937.6 \pm 0.1) + (-7.12 \pm 0.01) \Delta t$&
 $(732.7 \pm 0.1) + (2.90 \pm 0.02) \Delta t$\\
\hline
S100&15.4 &2005.42 & $(-980.5 \pm 0.1) + (-0.64 \pm 0.02) \Delta t$&
 $(554.8 \pm 0.1) + (-2.28 \pm 0.02) \Delta t$\\
\hline
S109& 17.3 &2009.43 & $(-863.8 \pm 0.2) + (6.44 \pm 0.06) \Delta t$&
 $(-797.0 \pm 0.2) + (-3.59 \pm 0.07) \Delta t$\\
\hline
S110& 16.9 &2009.43 & $(-790.8 \pm 0.1) + (-2.83 \pm 0.04) \Delta t$&
 $(-733.3 \pm 0.2) + (-1.28 \pm 0.05) \Delta t$\\
\hline
S146& 17.5 & 2004.38 &$(-1387.0 \pm 0.2) + (-4.77 \pm 0.05) \Delta t + (0.068 \pm 0.006) \Delta t^2$&
 $(-620.6 \pm 0.1) + (0.97 \pm 0.05) \Delta t + (0.035 \pm 0.006) \Delta t^2$\\
\hline
\end{tabular}
}
\end{table*}

\end{document}